\documentclass[12pt]{article}
\usepackage[margin=1in]{geometry}
\usepackage{amsmath,amsthm,amssymb}
\usepackage{bm}
\usepackage{graphicx}
\usepackage{tikz}
\usetikzlibrary{arrows.meta, positioning}
\usepackage[colorlinks=true, urlcolor=blue, linkcolor=blue, citecolor=magenta]{hyperref}
\usepackage{algorithm}
\usepackage{algpseudocode}
\usepackage{color}
\usepackage{comment}
\usepackage{enumerate}
\usepackage{multicol}
\numberwithin{equation}{section}
\usepackage{cleveref}

\newtheorem{theorem}{Theorem}[section]
\newtheorem{lemma}{Lemma}[section]
\newtheorem{corollary}{Corollary}[section]
\newtheorem{remark}{Remark}[section]
\newtheorem{proposition}{Proposition}[section]

\newtheorem{assumption}{Assumption}[section]
\theoremstyle{definition}

\begin{document}

\title{An iterative procedure for statistical calibration } 
\author{Haiyi Shi  \,and\, Don Estep} 
 \date{}

\maketitle 


\newcommand{\tui}{ \mathcal{U}(Q^{-1}(I_{i})\cap A)}
\newcommand{\ui}{\mathcal{U}(Q^{-1}(I_{i})) }
\newcommand{\E}{\mathbb{E}}
\newcommand{\Var}{\mathrm{Var}}
\newcommand{\MD}{\mathcal{D}}
\newcommand{\MC}{\mathcal{C}}
\newcommand{\MH}{\mathcal{H}}
\newcommand{\pMD}{\rho_{ \mathcal{D}}}
\newcommand{\pMDj}{\rho_{{\mathcal{D}}_j} }
\newcommand{\pMDk}{\rho_{{\mathcal{D}}_k} }
\newcommand{\PMDj}{{{P}}_{{\mathcal{D}}_j} }
\newcommand{\PMDk}{{{P}}_{{\mathcal{D}}_k} }

\newcommand{\PMD}{{{P}}_{{\mathcal{D}}} }

\newcommand{\pA}{\rho_{\Lambda}}
\newcommand{\Ql}{Q(\lambda)}
\newcommand{\Qjl}{Q_{j}(\lambda)}
\newcommand{\Qkl}{Q_{k}(\lambda)}
\newcommand{\Qjp}{{Q_j}_{\sharp}}
\newcommand{\Qkp}{{Q_k}_{\sharp}}
\newcommand{\Qy}{Q^{-1}(y)}
\newcommand{\Qjy}{Q^{-1}_{j}(y_j)}
\newcommand{\Qky}{Q^{-1}_{k}(y_j)}
\newcommand{\JQ}{\vert \mathrm{Jac}\, {Q}\vert}
\newcommand{\JQj}{\vert \mathrm{Jac}\, {Q_j}\vert}
\newcommand{\JQk}{\vert \mathrm{Jac}\, {Q_k}\vert}

\newcommand{\Dkl}{\mathrm{D_{KL}}}
\newcommand{\IL}{\int_{\Lambda}}
\newcommand{\ID}{\int_{\mathcal{D}}}
\newcommand{\IDj}{\int_{\mathcal{D}_j}}
\newcommand{\IDk}{\int_{\mathcal{D}_k}}

\newcommand{\barp}{\bar{\rho}}
\newcommand{\barP}{\bar{P}}
\newcommand{\hatP}{\hat{P}}
\newcommand{\hatp}{\hat{\rho}}

\abstract{
We study the iterative proportional fitting procedure (IPFP) under multiple pushforward constraints, motivated by model calibration. The goal is to find a distribution on model parameters whose pushforward by models matching data distributions.  We establish the convergence of IPFP under suitable regularity conditions and develop a variant of sequential importance sampling to approximate the iterative procedure. We also propose a related procedure for constructing a high-probability parameter region. Simulation studies illustrate the performance of these methods. Finally, we apply IPFP to COVID-19 ensemble forecasting. Variation of Covid strains across regions can lead to variation in surge rates and thus affect forecasting accuracy. To capture the variation, we estimate a probability distribution of epidemiological model parameters using early-stage surge data from an ensemble of regions. We then propagate the distribution through the model to forecast both the ensemble’s overall trend and its variability.
}

\section{Introduction}
The iterative proportional fitting procedure (IPFP) is initially motivated by survey statistics \cite{deming1940least}, where some population groups are under- or overrepresented in survey samples. IPFP reweighs these samples to match population proportions reported in census data. IPFP has applications in a wide range of fields. In economics, it appears as the RAS method for balancing input--output tables \cite{lahr2004biproportional}; in entropic optimal transport, it takes the form of the Sinkhorn algorithm \cite{cuturi2013sinkhorn}; more recently, in generative modeling, IPFP is applied to construct a shorter diffusion process which generates samples from the data distribution \cite{de2021diffusion}.
 
 All these applications require constructing a joint probability distribution from given marginals. To this end, IPFP formulates them as a constrained optimization problem; given a reference probability distribution $P_0$ and constraint sets $\{\MC_j\}_{j=1}^t$, find a probability distribution that minimizes the Kullback--Leibler (KL) divergence from $P_0$:
\begin{equation}
P^* = \underset{P \in \bigcap_{j=1}^{t} \MC_j}{\arg\min}
      \Dkl(P \Vert P_0),
\end{equation}
where $\Dkl$ denotes the KL-divergence. In the marginal-constrained setting, each $\MC_j$ is the set of probability distributions whose $j$-th coordinate marginal equals a given distribution, so their intersection consists of distributions satisfying all $t$ marginal constraints. 
When it exists, the minimizer $P^*$ is called the information projection (I-projection) of $P_0$ onto $\bigcap_{j=1}^{t} \MC_j$. The name reflects the fact that it minimizes the statistical distance KL-divergence to $P_0$, which is intuitively similar to the geometric property of projection operators in Hilbert space \cite{csiszar1975divergence}. 

IPFP approaches this multi-constraint optimization problem by decomposing it into $t$ single-constraint optimizations. It requires the solution to the individual constraint problem, that is, the I-projection onto each set $\MC_j$. Then, starting from $P_0$, IPFP successively computes the I-projection of the current distribution onto $\MC_1,\dots,\MC_t$ and repeats this cycle. Under suitable conditions, the iterates converge to $P^*$.

The convergence of this procedure has been studied in several settings; early work \cite{deming1940least,sinkhorn1967diagonal,kullback1968probability,bishop1969incomplete,fienberg1970iterative} focuses on finite discrete cases with given marginals constraints. Later, Csiszar \cite{csiszar1975divergence} shows the convergence for finite discrete probability measures onto intersection of linear sets, i.e., $P_1,P_2 \in \MC_j$ $\implies$ $\alpha P_1 + (1-\alpha)P_2 \in \MC_j $, for all $\alpha$ such that $\alpha P_1 + (1-\alpha)P_2$ is a probability measure. Dykstra \cite{dykstra1985iterative} developed a modified procedure for convex sets, which require only $\alpha \in [0,1]$.
For continuous distributions, Rüshendorff \cite{ruschendorf1995convergence} shows the convergence for bivariate densities given the marginal constraints. With the similar modification as Dykstra, Bhattacharya \cite{bhattacharya2006iterative} proves the convergence for probability measures onto intersection of convex sets using tools from Fenchel duality.  

Most applications of IPFP consider the marginal constraint, for which constructing the information projection is straightforward. This paper demonstrates the power of IPFP extends beyond these applications: we focus on a general case where the linear sets $\MC_j$ are given by pushforward constraints:
\begin{equation}\label{eq: constraint}
\MC_j = \{P \in \mathcal{P}(\Lambda) : \Qjp P = \PMDj \}, \quad j =1, \dots, t,
\end{equation}
where $P$ denotes a probability measure on the measurable space $\Lambda$, and each $Q_j$ is a measurable map that pushes $P$ forward to match a given measure $\PMDj$ on the measurable space $\MD_j$. 

 \textbf{IPFP for computer model calibration.} 
This multi-pushforward constraint problem is motivated by an important class of calibration problem, known as the statistical calibration problem \cite{bingham2024inverse,shi2025nonparametric}; in this setting, let $\Lambda \subseteq \mathbb{R}^l$ and $\MD_j \subseteq \mathbb{R}^{d_j}$, typically with $l \geq d_j$. 

The problem is to reconstruct an intrinsic \emph{random} and unobserved $\boldsymbol{\lambda} \in \mathbb{R}^l$ to match some observed \emph{random} $\{\bold{Y}_j \in \mathbb{R}^{d_j} \}_{j=1}^t $. Typically, $\bold{Y}_j$ and  $\boldsymbol{\lambda}$ have the following relationship:
\begin{equation}\label{eq: relation}
 \bold{Y}_j = Q_j(\boldsymbol{\lambda})  + \boldsymbol{\epsilon}, \,\, j=1,\dots, t,
\end{equation}
where the measurable function $Q_j : \mathbb{R}^l \to  \mathbb{R}^{d_j} $ represents the computer model and  $ \boldsymbol{\epsilon}$ is the measurement error and the index $j$ could represents varying experimental conditions, such as spatial locations or time points.

This formulation usually assumes that \emph{random variable} $\boldsymbol{\lambda}$ accounts for the most variability of $\bold{Y}$ while the noise $\boldsymbol{\epsilon}$  only plays a minor role. If not, we could treat $\boldsymbol{\epsilon}$ as a new random parameter of the model \cite{prasadan2026non,Butler_2020} or by deconvolution \cite{vadeboncoeur2026efficient}. Consequently, this paper focuses on
\begin{equation}\label{eq: relation2}
 \bold{Y}_j = {Q}_j(\boldsymbol{\lambda}), \,\, j=1,\dots, t.
\end{equation}
Since the pushforward constraint  \eqref{eq: constraint} is the distributional form of \eqref{eq: relation2}, we can solve the calibration problem by applying IPFP.

While this problem differs epistemologically from common calibration problems for finding a single best parameter to fit data driven mainly by measurement noise \cite{wilkinson2023introduction}, there is a fundamental necessity to model input parameters as random in many practical settings such as:

\begin{itemize}

\item Data is aggregated from experimental trials on a population of objects. Studying the variation of physical characteristics among these objects from data is crucial in various fields such as battery capacity \cite{baumhofer2014production,harris2017failure,dechent2021estimation}, cardiology \cite{lawson2018unlocking,britton2013experimentally}, cryo-electron microscopy \cite{penczek2006estimation,anden2018structural,espinosa2025cryo}, and neuroscience \cite{marder2011multiple}. For example, \cite{baumhofer2014production} analyzed the capacity fade of $48$ commercially available cells, which were all drawn from the same production lot. Even though cells were cycled under identical conditions, they exhibit a wide variability of capacity fade rate.


\item The calibration problem \eqref{eq: relation2} is often underdetermined; the model $Q$ is typically many-to-one, reflecting the loss of information in experiments. It is natural to quantify this indeterminacy by estimating a probability measure on all possible model inputs. This is justified by a nonparametric bayesian reasoning in \cite{bingham2024inverse,shi2025nonparametric}.  The Bayesian melding approach \cite{PR2000} and probabilistic approaches to inverse problem in geophysics \cite{Mosegaard1996,Mosegard2002} also addresses the indeterminacy similarly. Notably, these approaches differ from standard regularization, which forces a one-to-one mapping to produce a point estimate of model parameters.\cite{Stuart_Bayesian,bingham2024inverse}.
\end{itemize}

A major roadblock of early statistical calibration work \cite{bingham2024inverse,breidt2011measure} is that it only considers a single $Q: \mathbb{R}^{l} \to \mathbb{R}^{d} $ where $ d \leq l$; these work infers the distribution of $\boldsymbol{\lambda}$ from an estimate of $d$-dimensional probability distribution, e.g. kernel density estimation, from data. When $d$ is large, it can be extremely difficult to characterize the high-dimensional probability structure because of limited data size and the curse of dimensionality. In addition, the assumption $d \leq l$ can be restrictive in scenarios where observations are collected under many conditions but the dimensionality of model parameter $l$ remains small.

This paper extends early work from the single constraint to multiple pushforward constraints with $d_j \leq l, \,\, j =1, \dots, t$. This means we can select small values of  $d_j$ to bypass the curse of dimensionality. This does not necessarily result in an inaccurate solution, provided a sufficiently large number of constraints $t$ is utilized. Simulation study in subsection \ref{subsec: synthetic} demonstrates it is possible to reconstruct the covariance and bimodal structure from up to $12$ one-dimensional pushforward constraints. 
 
On the other hand, $\{Q_j\}_{j=1}^t$ doesn't require correlations between two different conditions $j_1$ and $j_2$ which are encoded by the joint $(Q_{j_1}, Q_{j_2})$. Therefore, this framework allows for learning a distribution from multiple and possibly unpaired datasets. An relevant application is the heterogeneity problem in cryo-electron microscopy \cite{penczek2006estimation,anden2018structural,espinosa2025cryo}.

\textbf{Related work on model calibration.}  
There is a scattering of work on calibrating parameter distributions given multiple datasets.
  \cite{chi2021sliced} proposed an iterative method different from IPFP. Roughly speaking, it computes information projections on different pushforward constraints and averages them to initialize the next iteration. However, its convergence is not known despite some empirical evidence. 
 
 Some work takes an optimization approach under a variational framework to define a solution to the problem \cite{Li2024inverse,espinosa2025cryo}. Several applied studies have proposed various sampling-based methods in neuroscience and cardiology \cite{marder2011multiple,britton2013experimentally,lawson2018unlocking}.
 
 More recently,  \cite{jiang2026iterative} also applies IPFP and attempt to show its convergence by constructing a product space, which reduces pushforward constraints into marginal ones. Specifically, \cite[thm 5.1]{jiang2026iterative} defines a joint mapping \[G := (Q_1, \dots, Q_t): \Lambda \subseteq \mathbb{R}^{l} \to \prod_{j=1}^t D_j \subseteq \mathbb{R}^{\sum d_j},\] such that $\Qjp P_{\Lambda}$ becomes the $j$-th marginal of the pushforward measure $G_{\sharp} P_{\Lambda}$. However, because $G_{\sharp} P_{\Lambda}$ generally is not absolutely continuous with product of its marginals (e.g. when $l < \sum d_j$), the convergence results for marginal constraints \cite{ruschendorf1995convergence} do not apply directly. In addition, even if it is absolutely continuous, the factorization of each marginal in \cite[thm 5.1, step 2]{jiang2026iterative} need more justification.

\textbf{Related convergence result.}  \cite{bhattacharya2006iterative} establishes convergence of a modified iterative I-projection algorithm in a much broader setting, including the pushforward constraint sets. Our convergence results complement \cite{bhattacharya2006iterative} by providing a direct convergence argument, with explicit I-projection and assumption on the reweighting factors, which are tailored to our scenario, and practical calibration applications.
 
\textbf{Relation to Schr\"odinger bridge.}  Suppose that, for every $\lambda\in\Lambda$, the map $s\mapsto Q(\lambda;s)$ induces a continuous path on $[0,T]$. Let $Q_j(\lambda)=Q(\lambda;s_j)$. Under a parameter distribution $P_{\Lambda}$, the time marginal of the induced stochastic process at $s_j$ is the pushforward $\Qjp P_{\Lambda}$. The multi-pushforward constraint problem therefore has a multi-marginal Schr\"odinger bridge interpretation: the corresponding path measure minimizes relative entropy with respect to the reference path measure induced by $Q$ and $P_0$, subject to the prescribed time marginals $\PMDj$ \cite{leonard2014survey}. IPFP on the parameter space $\Lambda$ then induces a corresponding iterative procedure on the path space. Our primary interest lies in calibration problems with $\Lambda\subseteq\mathbb{R}^l$. In this setting, the model-induced reference process need not be Markovian, and its joint output distribution may be singular with respect to the product of its marginals, as discussed above. Thus the setting differs from the nondegenerate diffusion references commonly used in Sch\"odinger bridge formulations \cite{pmlr-v206-stromme23a,dimarino2020optimal}.

\textbf{Summary.}  As a result, we believe this perspective of IPFP is theoretically interesting and useful in practical applications. We summarize three main contributions of this paper in the following:

 \begin{itemize}

 \item We prove the convergence of the IPFP under pushforward constraints \eqref{eq: constraint} given by measurable functions; We provide regularity conditions on the iterative sequence and alternative sufficient conditions only on the reference measure that ensure convergence.  
 \item  
For computer model calibration implementations, we propose a variation of sequential importance sampling method based on IPFP. We discuss its time complexity and establish its convergence under mild regularity condition. Simulation studies validate the algorithm's convergence and its robustness to misspecified reference measures. In addition, we also propose a relevant and practical procedure, which shares similar argument with simultaneous inference, for constructing a high-probability parameter region. 

 \item
We apply our procedure to COVID-19 ensemble forecasting, regarding 113 U.S. university counties with similar demographic and policy characteristics as an ensemble. We calibrate an SIR model using data from the early stages of the surge and forecast the ensemble’s future surge dynamics. This approach captures the overall trend while quantifying variation in surge rates across counties. Such variation might reflect the variation of Covid strains, and thus is important for understanding and forecasting surge dynamics.
 \end{itemize}
 
The remainder of the paper is organized as follows. Section \ref{subsec: pre} introduces notation and review basics of IPFP.  Section~\ref{sec: main} establishes the convergence of IPFP under pushforward constraints. Section~\ref{sec: alg} presents two algorithms: an IPFP algorithm for computer model calibration and a procedure for constructing a high-probability parameter region. Section~\ref{sec: app} illustrates these methods by one synthetic example and a real-data application to Covid-19 forecasting. Section \ref{sec: con} concludes this paper and discusses future directions.  For clarity, all technical proofs are deferred to Section \ref{sec: prof}.

\subsection{Preliminary}
\label{subsec: pre}

We start with notation and preliminary for results. Let $(\Lambda, \mathcal{F}_{\Lambda})$  be a measurable space and let $P$ and $P_0$ denote two probability measures on $(\Lambda, \mathcal{F}_{\Lambda})$. The KL-divergence of $P$ relative to $P_0$ is defined as
\[
\Dkl(P \| P_0) =
\begin{cases}
\displaystyle \int_{\Lambda} \ln \frac{dP}{dP_0} \, dP,
& \text{if } P \ll P_0, \\[6pt]
+\infty,
& \text{if } P \not\ll P_0.
\end{cases}
\]
For a given  $P_0$ and a set of probability measures $\MC$ on  $(\Lambda, \mathcal{F}_{\Lambda} )$, we are interested in finding a $P^*$ in $\MC$ that is ``closest'' to $P_0$ in the sense of KL divergence,
\begin{equation} 
\Dkl(P^* \Vert P_0) = \inf_{\bar{P} \in \MC} \Dkl(\bar{P} \Vert P_0) < +\infty.
\end{equation}
 This $P^*$ is called the information projection (I-projection) of $P_0$ onto $\MC$. Csiszár \cite{csiszar1975divergence} has shown that $P^*$ exists uniquely if the  $\MC$ is convex, variational-closed (i.e., closed in terms of total variation.) and there exists $\bar{P} \in \MC$ such that $\Dkl(\bar{P} \Vert P_0) < \infty$. Moreover, $P^*$ is the I-projection of $P_0$ onto the convex set $\MC$ if and only if it satisfies the triangle inequality: 
  \begin{equation}\label{eq: triangle}
 \Dkl(\bar{P} \Vert P_0) \geq \Dkl(\bar{P} \Vert P^*) + \Dkl(P^* \Vert P_0 ),
 \end{equation}
 for every $\bar{P} \in \MC $.

This paper studies the case where $\MC$ is given by pushforward constraints. For each $j = 1, \dots, t$, let $(\MD_j, \mathcal{F}_{\MD_j})$ be measurable spaces, and let $\PMDj$ be a probability measure on $(\MD_j, \mathcal{F}_{\MD_j})$. Given a measurable mapp $Q_j: (\Lambda, \mathcal{F}_{\Lambda}) \to (\MD_j, \mathcal{F}_{\MD_j})$, we define $\MC_j$ as the collection of all probability measures $P$ on $(\Lambda, \mathcal{F}_{\Lambda})$ whose pushforward measure under $Q_j$ is $\PMDj$, that is,
\begin{equation} 
\MC_j = \{P \in \mathcal{P}(\Lambda):  \Qjp P = \PMDj \}, \quad j =1, \dots, t,
\end{equation}
where $\mathcal{P}(\Lambda)$ denotes all the probability measures on measurable space $(\Lambda, \mathcal{F}_{\Lambda})$ and $\Qjp P $ denotes the pushforward measure of $P$ by $Q_j$ such that 
$\Qjp {P}(B) = {P}(Q_j^{-1}(B))$ for any measurable set $B \in \mathcal{F}_{\MD_j}$. Then, $\MC$ is defined as $\bigcap_{j=1}^t \MC_j$. It is straightforward to verify that $\MC_j$ and $\MC$ are both convex (moreover, they are linear set) and variational-closed.

To search for the I-projection of $P_0$ onto the intersection $\MC$, IPFP begins with an reference distribution $P_{1,0} = P_0$ on $\Lambda$ and sequentially computes the I-projection onto $\MC_1, \MC_2, \dots, \MC_t$, to obtain a sequence of distributions $P_{1,1}, P_{1,2}, \dots, P_{1,t}$. The process then enters its next cycle ($i=2$) by taking the final projection $P_{1,t}$ of the previous cycle as its new starting point, setting $P_{2,0} = P_{1,t}$. As a result, we obtain a sequence of probability measures $\{P_{i, j}: i \geq 1, j = 1,2, \dots t. \}$, where $i$ indexes the cycle and $j$ indicates the I-projection onto $\MC_j$. For ease of notation, we set $P_{i,t} = P_{i+1,0} $ for all $i$.

\section{The convergence of IPFP} \label{sec: main}
This section studies the convergence of IPFP for the multi-pushforward constraint problem:
\begin{equation}
\label{eq: pushP}
\min_{P \in \cap \MC_j} \Dkl(P \Vert P_0),
\qquad
{\MC}_j :=
\{P\in\mathcal P(\Lambda): \Qjp P =\PMDj \}.
\end{equation}

We begins with a sequence of probability measures  $\{P_{i,j}\}_{i,j}$ on $\Lambda$, which in the sequel is shown to be the I-projections for the iterative procedure. 
Given  $P_{1,0}$, $\{Q_j\}_{j=1}^{t}$, and probability measures $\{\PMDj\}_{j=1}^{t}$, for $ i \geq 1$, the sequence is given inductively by
\begin{alignat}{2}\label{eq: Piproj}
d P_{i,j}& = \frac{d \PMDj}{d (\Qjp P_{i,j-1})}(Q_j(\lambda))\, dP_{i,j-1} \quad & j &= 1,2,\dots,t \nonumber \\
P_{i+1,j}& = P_{i,t} \quad & j &= 0,
\end{alignat}
where $ \frac{d \PMDj}{d (\Qjp P_{i,j-1})}$ denotes the Radon-Nikodym derivative of $\PMDj$ w.r.t. $\Qjp P_{i,j-1}$.
 The absolute continuity condition, $\PMDj \ll \Qjp P_{i,j-1}$, is guaranteed by the following assumption:
\begin{assumption}
\label{assp: abscon}
There exists $\barP \in \MC = \cap_{k=1}^t \MC_k$ such that $\barP \ll P_{1,0}$.
\end{assumption}
The following lemma's proof is deferred in subsection \ref{proof: abscon}.
\begin{lemma}
\label{lem: abscon}
Under Assumption \ref{assp: abscon}, for every $i \geq 1$ and $t \geq j \geq 1$, $\barP \ll P_{i,j} $, and $\PMDk \ll \Qkp P_{i,j}$ for every $k=1, \dots, t$.
\end{lemma}
Furthermore, after the first cycle, $ i>1$, all probability measures $P_{i,j}$ are mutually absolute continuous.  The following proof is deferred in subsection \ref{proof: equivalence}.
\begin{corollary}
\label{cor: equivalence}
Under Assumption~\ref{assp: abscon}, for every $i\geq 2$ and
$1\leq j\leq t$,
\[
    P_{i,j}\sim P_{1,t},
\]
where $\sim$ denotes mutual absolute continuity.
\end{corollary}

Thus the construction~\eqref{eq: Piproj} is well-defined. Moreover, it yields probability measures that satisfy the pushforward constraints.
\begin{lemma}
\label{lem: changeov}
Under Assumption~\ref{assp: abscon}, for every $i \geq 1$, $t \geq j \geq1$, $P_{i,j} \in \MC_j$, that is, $\Qjp P_{i,j} = \PMDj$.
\end{lemma}
\begin{proof}
For every measurable set $B_j \subseteq \MD_j$, by the change of variable formula \cite{Bog07},
\begin{eqnarray}
P_{i,j}(Q^{-1}_{j}(B_j)) &=& \int  \frac{d \PMDj}{d {Q_j}_{\sharp} P_{i,j-1}} \big(Q_{j}(\lambda)\big) {\chi_{B_j}} (Q_{j}(\lambda)) \, dP_{i,j-1} \nonumber \\
&=& \int_{B_j} \frac{d \PMDj}{d {Q_j}_{\sharp} P_{i,j-1}}  \, d (\Qjp P_{i,j-1}) = \PMDj(B_j).
\end{eqnarray}
If $B_j = \MD_j$ in the above equality, $P_{i,j}(\Lambda) = 1$.
\end{proof}

Inductively, we write $P_{n,r} (n \geq 1, t \geq r \geq 1)$ from $P_{1,0}$
\begin{eqnarray}\label{eq: formula}
d P_{n,r} &=&  \prod_{1\leq i\leq n} \prod_{1\leq j \leq j_i} \frac{d \PMDj}{d \Qjp P_{i,j-1}}(\Qjl) \, dP_{1,0} := \prod_{1\leq i\leq n} \prod_{1\leq j \leq j_i} w_{i,j}(\Qjl) \, dP_{1,0} \nonumber \\
 &:= & W_{n,r}(\lambda)\, dP_{1,0}, 
\end{eqnarray}
where $j_i = t$ if $i < n$, and $j_i = r$ if $i = n$, $w_{i,j}$ can be regarded as the reweight term at each iteration, and $W_{n,r}$ as the total reweight term. 
This product form allows $(\ln \frac{dP_{n,r}}{d P_{1,0}})$ to factorize, leading to pythagorean-type law of the sequence in the sense of KL-divergence in Lemma~\ref{lem: py}. To show this, we first examine if the integral $\Dkl(P_{i,j}|| P_{i,j-1})$ is well-defined. The following proof is deferred in subsection \ref{proof: anyP}.

\begin{lemma} \label{lem: anyP}
Under Assumption \ref{assp: abscon}, for any $\barP_j \in \MC_j$, the following integrals are well-defined and
\begin{equation}
\label{eq: anyP}
\IL  \ln \frac{dP_{i,j}}{dP_{i,j-1}} \, d \barP_j =\Dkl(\PMDj \Vert \Qjp P_{i,j-1})= \Dkl(P_{i,j}\Vert P_{i,j-1}).
\end{equation}
\end{lemma}

The I-projection also requires $ \Dkl(P_{i,j}\Vert P_{i,j-1}) $ is finite for all $i, j$. This follows from the following assumption. Its proof is deferred in subsection \ref{proof: py}.
 \begin{assumption}\label{assp: exist}
$\inf_{\barP \in \bigcap_{j=1}^{t} \MC_j } \Dkl (\barP \Vert P_{1,0}) < \infty $.
\end{assumption}

\begin{lemma}[Pythagorean law] \label{lem: py}
Under Assumption \ref{assp: exist}, 
\begin{equation}
\label{eq: py}
\Dkl(\barP \Vert P_{1,0}) = \Dkl(\barP \Vert P_{n,r}) + \sum_{i=1}^{n}\sum_{j=1}^{j_i} \Dkl(P_{i,j}\Vert P_{i,j-1}),
\end{equation}
for any $n, r \geq 1$ and any $\barP \in \bigcap_{j=1}^{t} \MC_j $ such that $\Dkl(\barP \Vert P_{1,0}) < \infty$.
\end{lemma}

It is related to Pythagorean law for the I-projection proven in \cite{csiszar1975divergence} and when $\Dkl(P_{i,j}\Vert P_{i,j-1})$ is finite, consequently we have:
 \begin{theorem} \label{thm: Iproj}
Under Assumption \ref{assp: exist}, for every $i \geq 1$ and $t \geq j \geq 1$, $P_{i,j}$ is the unique I-projection of $P_{i,j-1}$ onto $\MC_j$. 
\end{theorem}

\begin{proof}
First, Lemma \ref{lem: changeov} shows that $\Qjp P_{i,j} = \PMDj$, so $P_{i,j} \in \MC_j$. 

Second, let $\barP_j \in \MC_j$ be any probability measure. If  $\Dkl(\barP_j \Vert P_{i,j-1}) = \infty$, the inequality $\Dkl(\barP_j \Vert P_{i,j-1}) > \Dkl(P_{i,j} \Vert P_{i,j-1})$ holds trivially, because $\Dkl(P_{i,j} \Vert P_{i,j-1}) < \infty$ by Assumption \ref{assp: exist}.

If $\Dkl(\barP_j \Vert P_{i,j-1}) < \infty$, then $\barP_j \ll P_{i,j-1}$. The construction of $P_{i,j}$ by \eqref{eq: Piproj} also implies $\barP_j \ll P_{i,j}$. (since any set where $\frac{d\PMDj}{d\Qjp P_{i,j-1}}$ is zero must also have $\barP_j$-measure zero by the pushforward constraint). Thus,
\begin{eqnarray}
\Dkl(\barP_j \Vert P_{i,j-1}) &=& \IL \ln \frac{d\barP_j}{dP_{i,j-1}} \, d\barP_j \nonumber \\
&=& \IL \ln \left( \frac{d\barP_j}{dP_{i,j}} \frac{dP_{i,j}}{dP_{i,j-1}} \right) \, d\barP_j \nonumber \\
&=& \Dkl(\barP_j \Vert P_{i,j}) + \IL \ln \frac{dP_{i,j}}{dP_{i,j-1}} \, d\barP_j \nonumber \\
&\stackrel{\eqref{eq: anyP}}{=}&  \Dkl(\barP_j \Vert P_{i,j}) +  \Dkl(P_{i,j} \Vert P_{i,j-1}).
\end{eqnarray}
Because $\Dkl(\barP_j \Vert P_{i,j}) \geq 0$, we have $\Dkl(\barP_j \Vert P_{i,j-1}) \geq \Dkl(P_{i,j} \Vert P_{i,j-1})$, where the equality holds if and only if $\Dkl(\barP_j \Vert P_{i,j}) = 0$, that is, $\barP_j = P_{i,j}$. 
\end{proof}
Now we obtain the basic property of the iterative procedure.
\begin{proposition} \label{prop: basic} 
Let $\Vert \cdot \Vert_{TV}$ denote the total variation norm for measures. Under Assumption \ref{assp: exist}, we have: \newline
1. $\Dkl(P_{n,j} \Vert P_{n,j-1}) \to 0$ and $\Vert P_{n,j} - P_{n,k} \Vert_{TV} \to 0 $ uniformly with respect to $j,k$, as $n\to \infty$. \newline
2. $\Vert \Qjp P_{n,k} - \PMDj \Vert_{TV} \to 0$ uniformly with respect to $k,j$, as $n \to \infty$.
\end{proposition}

The convergence of the sequence $\{\sum_{i=1}^{n}\sum_{j=1}^{j_i} \Dkl(P_{i,j}\Vert P_{i,j-1})\}_{n}$ does not imply convergence of $\{\sum_{i=1}^{n}\sum_{j=1}^{j_i}\Vert P_{i,j} - P_{i,j-1} \Vert_{TV}\}_{n}$, and hence convergence of $P_{n,j}$. Intuitively, if $\Dkl(P_{n,j} \Vert P_{n,j-1}) \sim \mathcal{O}({{n}^{-2- \epsilon}}) $ for some $\epsilon > 0$ so that by the Pinsker's inequality,  ${\Vert P_{n,j} - P_{n,j-1} \Vert_{TV}} \sim \mathcal{O}({{n}^{-1- \frac{\epsilon}{2}}}) $ to ensure the convergence of $P_{n,j}$.
Thus, to have $P_{n,j}$ converge to a limit in $\bigcap_{j=1}^{t} \MC_j$, we need stronger regularity conditions on the reweighting factors.

\begin{assumption}\label{assp: boundness}
For every $ t \geq j \geq 1$, $\{\frac{d \PMDj }{d \Qjp P_{i,j-1}}\}_{i\geq1}$ is uniformly bounded above $\PMDj$-a.e..
\end{assumption}

\begin{assumption} \label{assp: u.i.}
Assume the sequence $\{ \sum_{i=1}^n \ln \frac{d \PMDj}{d \Qjp P_{i, j-1}}  \}_{n>1}$ is uniformly integrable in $L^1(\PMDj)$, for every $ t \geq j \geq 1$.
\end{assumption}
The following lemmas' proof is deferred in subsection \ref{proof: limequal} and \ref{proof: wisu.i.}.
\begin{lemma}\label{lem: limequal}
Under Assumption \ref{assp: exist}, \ref{assp: boundness} and \ref{assp: u.i.}, for every $t\geq r\geq1$,
\begin{equation}
\lim_{n \to \infty} \Dkl(P_{n,r} \Vert P_{1,0}) = \sum_{i=1}^{\infty}\sum_{j=1}^{t} \Dkl(P_{i,j}\Vert P_{i,j-1}).
\end{equation}
\end{lemma}

\begin{lemma}\label{lem: wisu.i.}
Under Assumption \ref{assp: exist}, \ref{assp: boundness} and \ref{assp: u.i.}, $\{W_{n,r}\}_{n \geq 1}$ is uniformly integrable w.r.t. $P_{1,0}$, where $\{W_{n,r}\}$ is defined in~\eqref{eq: formula}.
\end{lemma}

The following convergence theorem follows the strategy in \cite[Theorem 3.1]{ruschendorf1995convergence}, and its proof is deffered in Subsection~\ref{proof: converge}.
\begin{theorem}\label{thm: converge}
Let $P^*$ be the unique I-projection of $P_{1,0}$ onto $\MC = \bigcap_{j=1}^t \MC_j$. Under Assumption  \ref{assp: exist}, \ref{assp: boundness} and \ref{assp: u.i.}, we have:
\begin{enumerate}
    \item $\Dkl(P^* \Vert P_{n,j}) \to 0$ as $n \to \infty$.
    \item $\Vert P_{n,j} - P^* \Vert_{TV} \to 0$ as $n \to \infty$.
\end{enumerate}
\end{theorem}

Assumptions \ref{assp: boundness} and \ref{assp: u.i.} are abstract and are stated in terms of the whole sequence. Thus, we work out conditions imposed on the first a few iterates of the sequence instead.
We first assume 
\begin{assumption} \label{assp: mutual}
$\PMDj \sim \Qjp P_{1,0}$ for every $j$.
\end{assumption}
 By Corollary~\ref{cor: equivalence},  the pushforward of $P_{i,j}$ and $\PMDj$ are mutually absolute continuous for every $j$ after first cycle. Thus if we reindex $P_{1,0}$ after first cycle, the above assumption is automatic.
 The following assumption quantifies the mismatch between $\Qjp P_{1,0}$ and $\PMDj$ by the essential oscillation of $\ln \frac{d \PMDj}{d \Qjp P_{1,0}}$.
\begin{assumption}
\label{assp: mismatch}
For a bounded measurable function $h$ on $\MD_j$, define
$$
  \operatorname{osc}_{\PMDj}(h)
  :=
  \operatorname*{ess\,sup}_{\PMDj}h - \operatorname*{ess\,inf}_{\PMDj}h.
$$
Assume there exists a constant $\eta > 0$ such that for every $t \geq j \geq 1$,
 $$
    \ln\frac{d\PMDj}{d\Qjp P_{1,0}}
    \in L^\infty (\PMDj),
    \qquad
    \operatorname{osc}_{\PMDj}
    \left(\ln\frac{d\PMDj}{d\Qjp P_{1,0}}\right)
    \leq\eta.
  $$
\end{assumption}
The following assumption quantifies the weak dependence between $\{Q_j\}_{j=1}^t$ under $P_{1,0}$.
\begin{assumption}
\label{assp: weak}
For each $j$, let $\MD_j$ be a standard Borel space and define a regular conditional probability as
$$
  K_j(y_j,A)
  :=
  P_{1,0}\bigl((Q_k)_{k\ne j}\in A\mid Q_j=y_j\bigr),
  \qquad A\subseteq\prod_{k\ne j}\MD_k\ \text{measurable},
$$
Thus $K_j(y_j,\cdot)$
is the conditional distribution of all the other
outputs when $Q_j=y_j$ under $P_{1,0}$. Assume there exists a constant $\delta\in[0,1]$ such that,
for every $1\leq j\leq t$,
\begin{equation}\label{eq: wd-kernel}
  \frac12
  \|K_j(y_j,\cdot)-K_j(y_j',\cdot)\|_{TV}
  \leq\delta
\end{equation}
for $(\PMDj\times\PMDj)$-almost every $(y_j,y_j')$.\end{assumption}

\begin{proposition}
\label{prop: weak-dependence-bounds}
Under Assumption \ref{assp: mutual}, \ref{assp: mismatch} and \ref{assp: weak}, assume there exists $B>\eta$ such that   \begin{equation}
   \delta \leq \frac{e^{B-\eta}-1}{e^{(t-1)B} -1}
    \label{eq: wd-invariant-bound}
  \end{equation}
Then, for every $n\geq1$ and
$1\leq j\leq t$,
\begin{equation}
  \operatorname{osc}_{\PMDj}
  \left(\sum_{i=1}^{n}\ln \frac{d \PMDj}{\Qjp P_{i,j-1}} \right)\leq B,
  \quad
  e^{-2B}\leq \frac{d \PMDj}{\Qjp P_{n,j-1}} \leq e^{2B},
  \quad \PMDj\text{-a.e.}
  \label{eq: wd-conclusions}
\end{equation}
In particular, Assumption~\ref{assp: boundness} holds.
If Assumption~\ref{assp: exist} also holds, then
Assumption~\ref{assp: u.i.} holds, and Theorem~\ref{thm: converge} follows.
\end{proposition}
The proof of proposition \ref{prop: weak-dependence-bounds} is deferred in Subsection~\ref{proof: weak-dependence-bounds}.

\section{Algorithms} \label{sec: alg}
For implementations of computer model calibrations, we present two algorithms: the first one for approximating a density solution by the iterative procedure, and  the second one for recovering a high-probability region.  
The first algorithm can be considered as a variation of sequential importance sampling. Intuitively, \eqref{eq: formula} suggests we can sequentially reweigh samples from the reference distribution $P_{1,0}$ at each iteration to obtain the I-projection. To implement this, set  $\Lambda \subseteq \mathbb{R}^{l}$ and $\MD_j \subseteq  \mathbb{R}^{d_j}$, initialize a set of Latin hypercube samples $\{\lambda^s\}_{s=1}^S$ on $\Lambda$ with normalized probability weight $\{W_{1,0}^s\}_{s=1}^S$. We partition $\MD_j$ into disjoint cells $ \MD_j = \cup_{m=1}^{M_j}{\Delta_{j}^{m}}$ for each $j =1, \dots, t.$ Then, the j-th partition induces a partition of $\{1,2,\dots, S\}$ by $\Lambda_{j}^m := \{ s:  Q_j(\lambda^s) \in \Delta^m_{j}  \}$. Each reweight term, $w_{i,j}^s$, is iteratively computed as
\begin{equation} \label{eq: weight}
w_{i,j}^s = \sum_{m=1}^{M_j}  \frac{P_{\MD_j}(\Delta_j^{m})}{\sum_{u \in \Lambda_{j}^{m} } W_{i,j-1}^u }\, \chi_{\Lambda_{j}^m} (s),
\end{equation}
where $W^u_{i,j-1}$ is the weight for sample $u$ at $(i, j-1)$ iteration.
We apply this $w^s_{i,j}$ to obtain the updated weight $W_{i,j}^s = W_{i,j-1}^s \, w^s_{i,j}$. The procedure is repeated for every $Q_j$ until $\sum_{s=1}^{S} \sum_{j=1}^{t} |W_{i,j}^s -W^s_{i,j-1}| \leq \epsilon $ for some threshold $\epsilon >0$. The complete procedure is shown in Algorithm \ref{alg: iterative}. 

\begin{algorithm}
\caption{Sequential importance sampling }\label{alg: iterative}
\begin{algorithmic}[1] 
\Require Computer codes $\{Q_j\}$; Data distributions $\{P_{\MD_j}\}$.
\Require Samples $\{\lambda^s\}_{s=1}^{S} \in \Lambda$ with normalized probability weight $\{W^{s}\}_{s=1}^{S}$;  
\Require Maximum iteration $N$; Tolerance $\epsilon$.
\State Discretize $\MD_j$ into $M_j$ cells: $\{\Delta_j^{m}\}_{m=1}^{M_j} $ for $j =1,2, \dots, t$; 
\State  Compute the probability of each cell $\{P_j^{m} \}_{m=1}^{M_j}$, such that $P_j^{m} = \PMDj (\Delta_j^{m})$;
\State Define $\Lambda_{j}^m := \{ s:  Q_j(\lambda^s) \in \Delta^m_{j}  \},$ $j = 1,2, \dots, t$ and $ m = 1,2, \dots M_j$.
  \For{(cycle index) $ i = 1,2, \dots, N: $ } 
  \State $\eta \gets 0$;
    \For{(projection index) $j = 1, 2, \dots, t: $ }
    	\For{(Cell index) $m = 1,2, \dots, M_j: $ } \Comment This loop can be done in parallel. 
     \State  $\sharp P \gets \sum_{\lambda^s \in \Lambda_j^m} W^s$;
     \State  $w \gets {P_j^m}/{(\sharp{P})}$;
     \State $\eta \gets \eta + |1- w| \sum_{\lambda^s \in \Lambda^{m}_{j}} W^s $;
     \State $W^s \gets W^s  w$ for $s \in \Lambda^{m}_{j}$;
        		\EndFor
                \EndFor
                \If{$ \eta \leq \epsilon$ }
	 Break;
	\EndIf
        \EndFor
   \State \Return $\{W^s\}^{S}_{s=1}$

\end{algorithmic}
\end{algorithm}

\begin{remark}
The iteration process has a complexity of $\mathcal{O}\big(N (t S + \sum_{j=1}^{t}M_j)\big)$, with $N$, $t$, and $S$ corresponding to the number of cycles, computer models, and the parameter sample size on $\Lambda$, respectively. In addition, a major factor of the algorithm's total complexity is the evaluation time of the computer models at parameter samples with size $S$.
\end{remark}
\begin{remark}
Estimating distributions in high-dimensional spaces is extremely difficult, as the curse of dimensionality means the need for prohibitively large sample sizes.
So, we mainly consider low-dimensional data, for which discretization of $\MD_j$ into cells and the computation of their respective probabilities in algorithm \ref{alg: iterative} is feasible.
 \end{remark}
Because each individual weight $w_{i,j}(Q_j)$ in \eqref{eq: formula} is $\sigma(Q_j)$-measurable, the total reweighting term is measurable with respect to the sub $\sigma$-algebra $\sigma(Q_j: j=1, \dots, t)$. Consequently, the iterative process only modifies probabilities at the resolution of this sub $\sigma$-algebra. More precisely, the finite partition of $\MD_j$ generates a finite sub $\sigma$-algebra, $\mathcal{G}_j := \sigma(\Delta_j^m: m =1,\dots, M_j)$, so the iteration computes probabilities on sets within:$$\mathcal{G}_{\Lambda} := \sigma(Q_j^{-1}(\Delta^m_{j}): m = 1, \dots, M_j, j=1, \dots, t).$$Therefore, the weighted samples $W_{i,j}^s$ induce a probability measure on the measurable space $(\Lambda, \mathcal{G}_{\Lambda})$, defined as 
$$
P_{i,j}(A) = \sum_{\lambda^s \in A} W_{i,j}^s, \,  \forall A \in \mathcal{G}_{\Lambda}.
$$ 
Thus, the pushforward constraint for every $j$ is:
$$
\MC_j = \{ P \, \text{on} \, (\Lambda, \mathcal{G}_{\Lambda}) :
P(Q_j^{-1}(\Delta^{m}_j)) = \PMDj(\Delta^{m}_j)) \, \text{for every}\, m
\}
$$

Since this is a finite case, the convergence of the sequence $\{P_{i,j}\}$ defined by the Algorithm \ref{alg: iterative} follows from \cite[Theorem 3.2]{csiszar1975divergence}.
\begin{theorem}
Let $P^*$ be the I-projection of $P_{1,0}$ onto $ \cap_{j=1}^{t} \MC_j$. Under the assumption~\ref{assp: abscon}, $\Vert P_{n,j} - P^*\Vert_{TV} \to 0$.
\end{theorem}

\subsection{Constructing a high probability parameter region}

The iterative procedure seeks a parameter distribution whose pushforwards match the observed data distributions. A less ambitious but very practical goal is to recover a high probability region of the parameter from data. Given $t$ data distributions $\{\PMDj\}_{j=1}^t$ and corresponding computer models $\{Q_j\}_{j=1}^{t}$, for which $\Qjp P_{\Lambda} = \PMDj$, the goal is to find a set $A$ in $\Lambda$ such that $$P_{\Lambda}(A) \geq 1 - \delta,$$
where $P_{\Lambda}$ is the unknown true parameter distribution, and $ \delta \in [0,1)$ is reasonably small. Our approach to this problem is similar to simultaneous inference: using a union bound to control whether any component guarantee fails. Define $A = \bigcap_{j=1}^t Q_j^{-1}(I_j)$, where $I_j \subseteq \MD_j$. Then,
$$
P_{\Lambda}(A^c) \leq \sum_{j=1}^t P_{\Lambda}(Q_j^{-1}(I_j^c)) = \sum_{j=1}^t \PMDj(I_j^c) < \delta,
$$
provided $\PMDj (I_j) \geq 1- \frac{\delta}{t}$.

To estimate each $\PMDj(I_j)$ from data, we can construct some tolerance region $\hat{I}_j$ such that it covers at least a proportion $1-{\delta}/{t}$ of $\PMDj$ with a confidence level of $1 - \frac{1-\beta}{t}$, that is,
$$
P_{\text{data}}( E_j) \geq 1 - \frac{1-\beta}{t}, \qquad E_j := \{ \PMDj(\hat{I}_j) \geq 1- {\delta}/{t} \}.
$$ 
This implies 
$$
P_{\text{data}}( \bigcap_{j=1}^t E_j) \geq \beta.
$$
Consequently, if $\hat{A} = \bigcap_{j=1}^{t} Q_j^{-1}(I_j)$, then 
$$
P_{\text{data}} ( P_{\Lambda} (\hat{A}) \geq 1- \delta ) \geq \beta.
$$
The tolerance regions are usually constructed using order statistics, see some results and packages in \cite{david2004order, tol2001, young2010tolerance}.
The folllowing algorithm \ref{alg: domain} approximates a high probability region in $\Lambda$ from tolerance interval in $\MD_j$ by rejection sampling.
\begin{algorithm}
\caption{sampling a high-probability parameter region}
\label{alg: domain}
\begin{algorithmic}[1]
\Require Models $Q_j:\Lambda\to\MD_j$, $j=1,\ldots,t$.
\Require Datasets $D_j$, $j=1,\ldots,t$.
\Require Parameter samples
    $\{\lambda^s\}_{s=1}^S\subseteq\Lambda$.
\Require Target probability $1-\delta$ and confidence $\beta$,
    where $\delta,\beta\in(0,1)$.
\For{$j=1,\ldots,t$}
    \State Construct a tolerance region $\widehat I_j$
        satisfying
        $P_{\mathrm{data}}
        \bigl(\PMDj(\widehat I_j)\ge  1-\frac{\delta}{t}  \bigr)\ge 1-\frac{1-\beta}{t} $.
\EndFor
\State $S_I\gets\{1,\ldots,S\}$.
\For{$j=1,\ldots,t$}
    \State $S_I\gets
        \{s\in S_I:Q_j(\lambda^s)\in\widehat I_j\}$.
        \Comment{This loop can be done in parallel}
    \If{$S_I=\varnothing$}
        \State \textbf{break}
    \EndIf
\EndFor

\State \Return $\{\lambda^s:s\in S_I\}$.
\end{algorithmic}
\end{algorithm}

\section{Applications} \label{sec: app}
\subsection{A synthetic example} \label{subsec: synthetic}
We consider the model of radioactive decay defined as
$Q(x,y;\theta) = x e^{-\theta y}$, where $x,y$ are the parameters and  $\theta$ is the control variable. We consider $12$ different $\theta \in \{0.01,0.05,0.1,0.2,0.3,$ $0.4,0.5,0.6,0.7,0.8,0.9,1.0 \}$, leading to $Q_1, \dots, Q_{12}$.
 We assume the true parameter distribution, $P_{\Lambda}$, 
is a two-dimensional Gaussian mixture distribution
 $$ ({X}, {Y}) \sim 0.7 \,  \mathcal{N}( \mu_1,\Sigma_1 ) + 0.3 \, \mathcal{N}( \mu_2, \Sigma_2 ).$$
Here, the means and covariance matrices are $\mu_1 = (-1,-1)$ with $\Sigma_1 = \big(\begin{smallmatrix} 0.5 & 0\\ 0 & 0.5 \end{smallmatrix}\big)$, and $\mu_2 = (1,1)$ with $\Sigma_2 = \big(\begin{smallmatrix} 1 & 0.7\\ 0.7 & 1 \end{smallmatrix}\big)$.
From this true distribution $P_{\Lambda}$, 10,000 parameter samples are drawn. These samples are then passed through each model $Q_j$ to generate a one-dimensional simulated dataset.
Now we apply the sequential importance sampling to recover $P_{\Lambda}$ and compare solutions given different priors and models.

\begin{figure}[htbp]    \centering
    
    \includegraphics[width=0.32\textwidth]{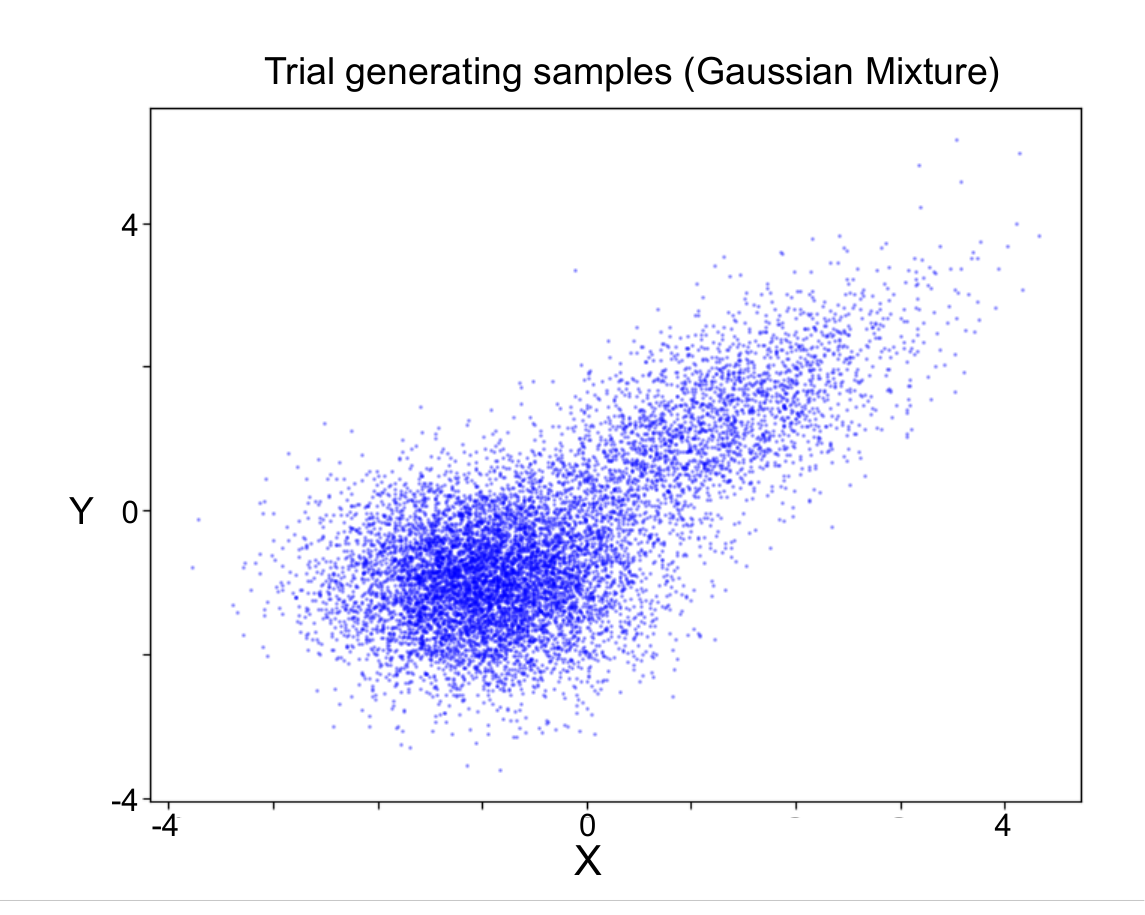}
    \hfill
    \includegraphics[width=0.32\textwidth]{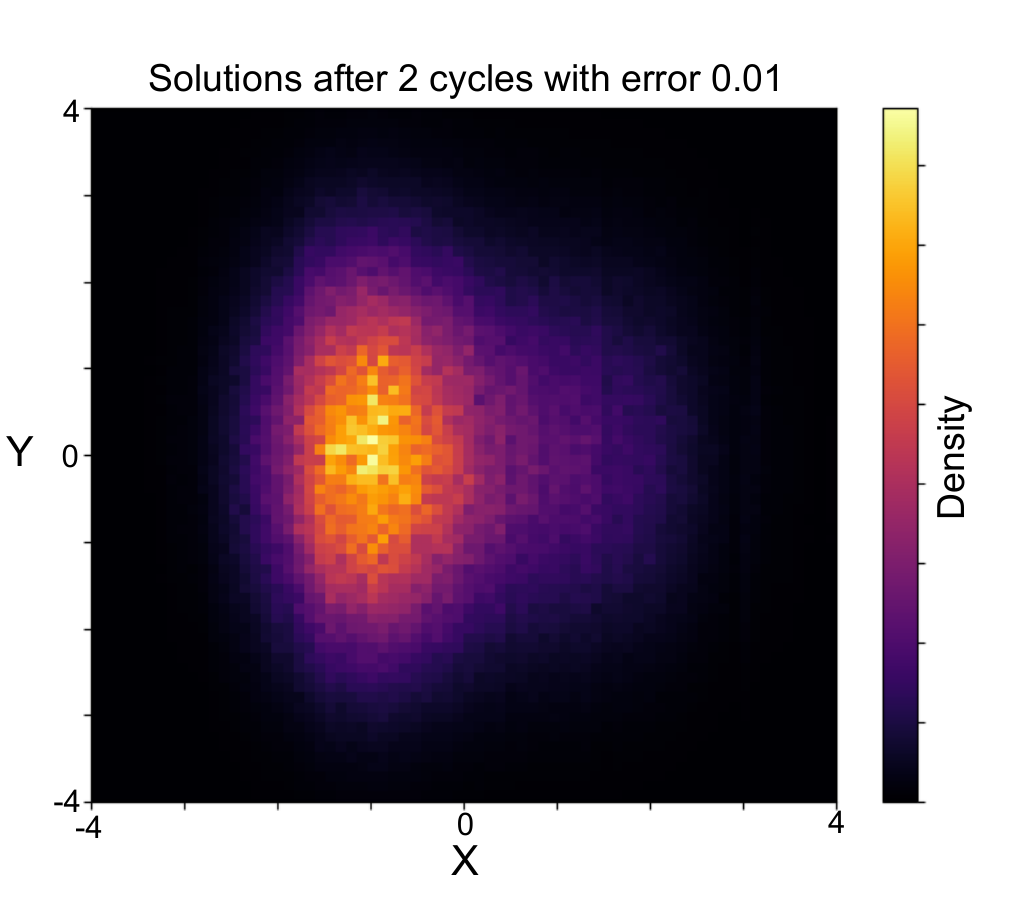}
    \hfill
    \includegraphics[width=0.32\textwidth]{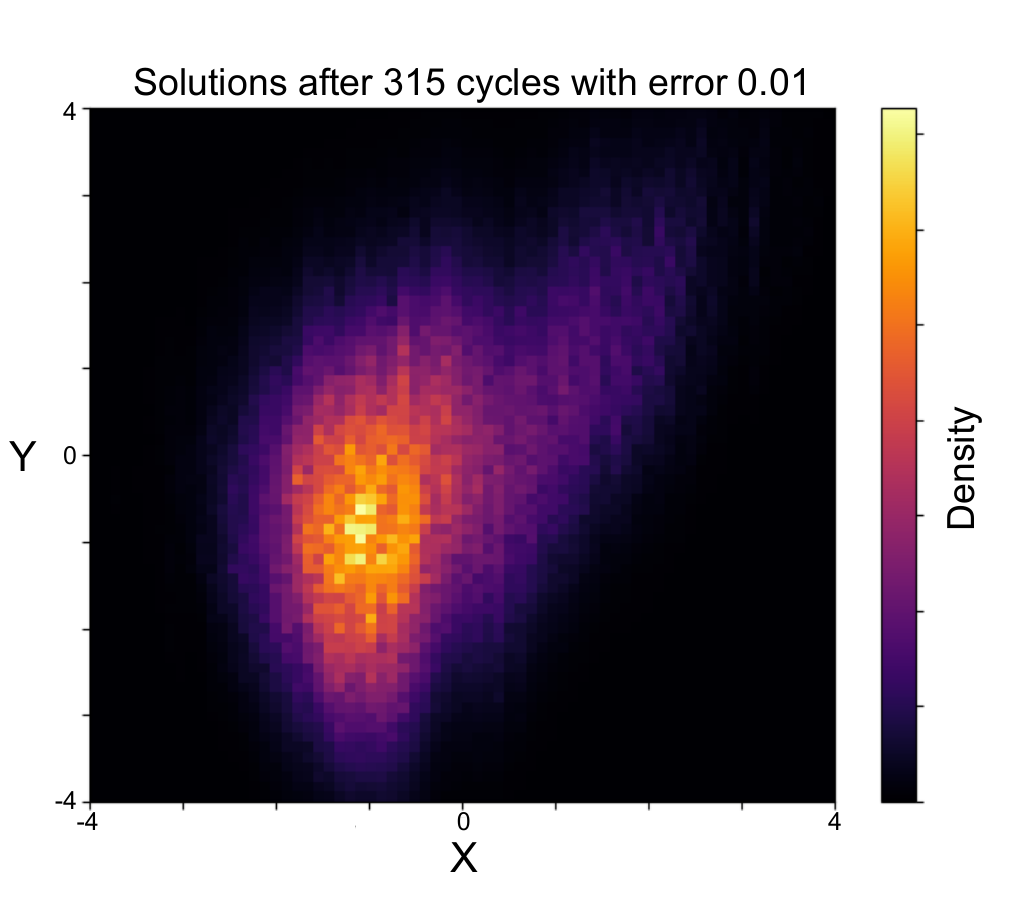}
    
    \vspace{1em} 
    
    \includegraphics[width=0.32\textwidth]{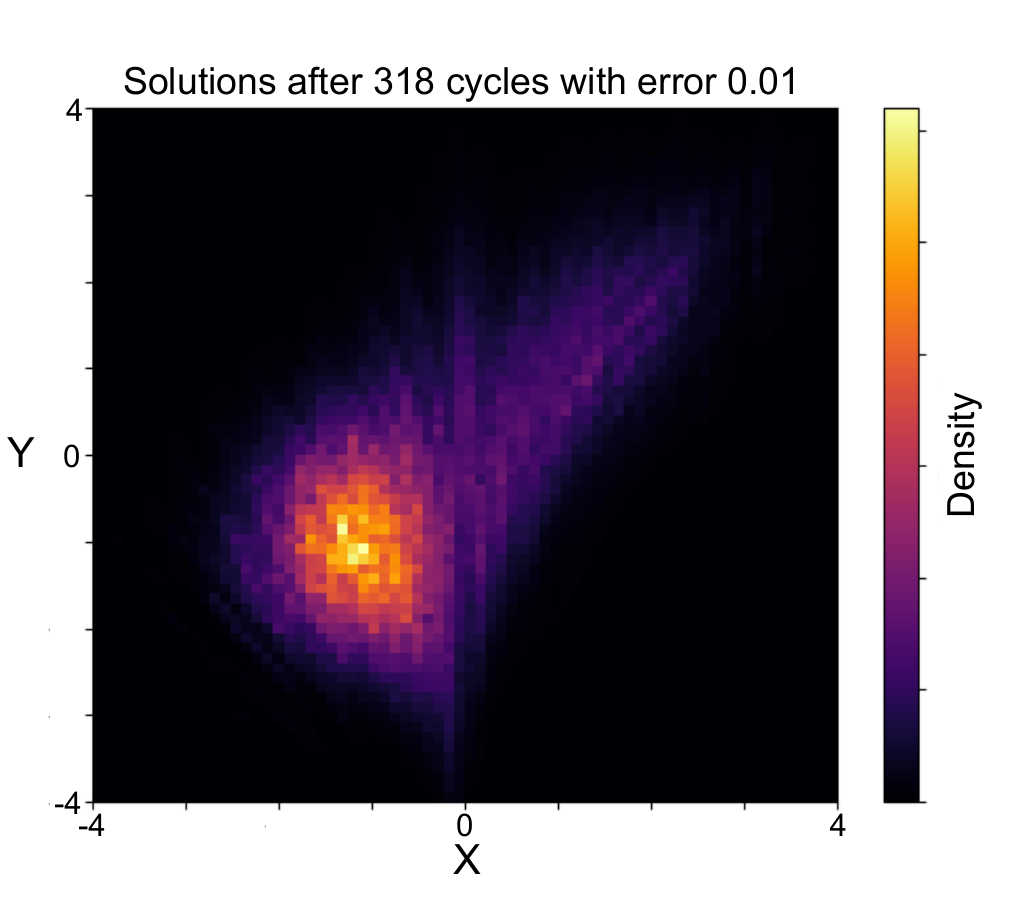}
    \hfill
    \includegraphics[width=0.32\textwidth]{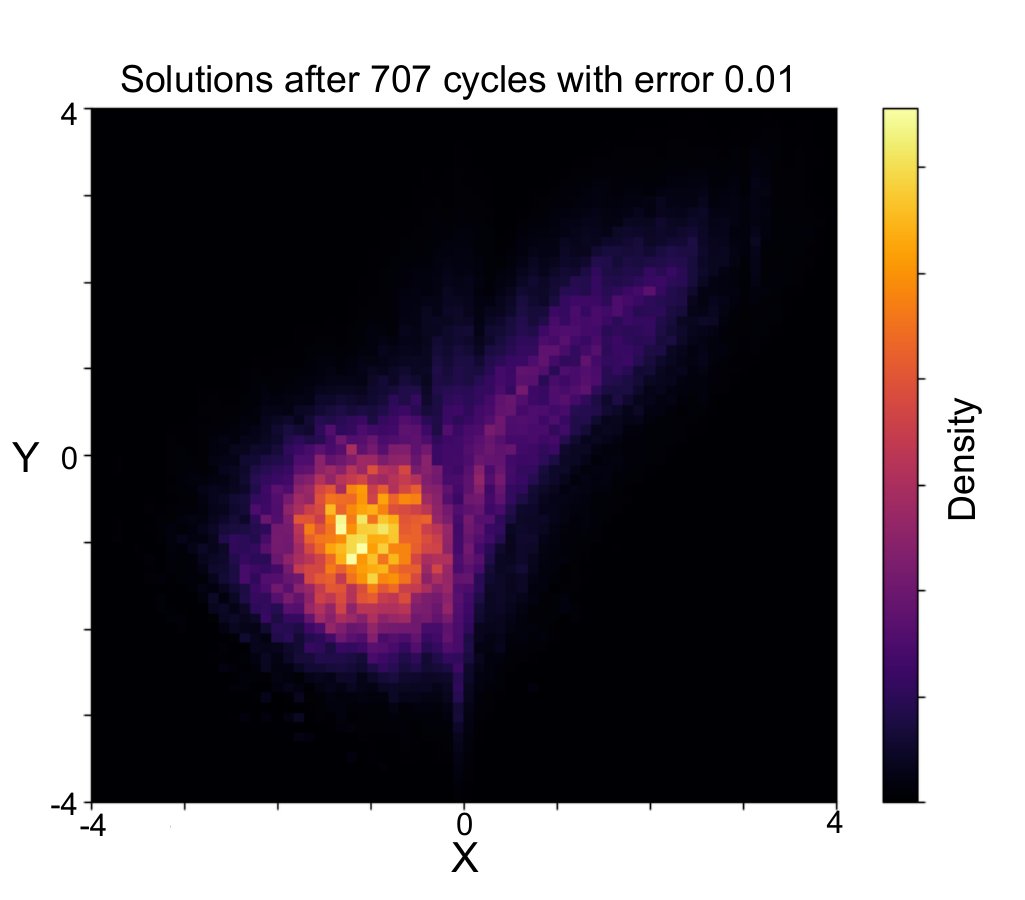}
    \hfill
    \includegraphics[width=0.32\textwidth]{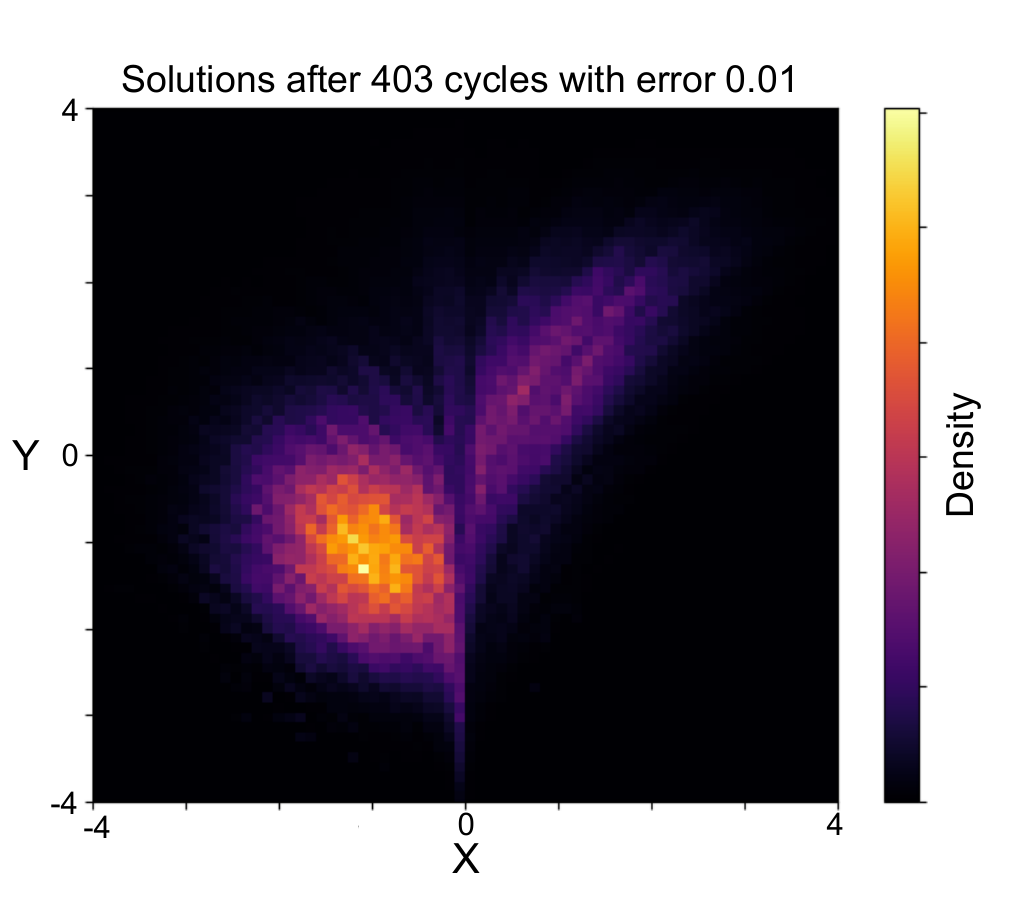}
    
    \caption{first row: $10000$ trial generating samples, converged solution by $Q_1$, and converged solution by $\{Q_1, Q_2, Q_3\}$. Second row: three converged solutions by $\{Q_1, \dots, Q_7\}$, $\{Q_1, \dots Q_{12} \}$, and $\{Q_6, \dots, Q_{12}\}$.}
    \label{fig: compare1}
\end{figure}

In the first scenario,  we assume $ P_{1,0} = \mathcal{N}\big((0,0), \big(\begin{smallmatrix}
  2 & 0\\
  0 & 2
\end{smallmatrix}\big) \big) $ and compute solutions using some subsets of models, $\{Q_1\}$, $\{Q_1,Q_2, Q_3\}$ $\{Q_1 \dots, Q_7\}$, $\{Q_1, \dots, Q_{12}\}$, and $\{Q_6, \dots, Q_{12}\}$. The results, shown in Figure \ref{fig: compare1}, demonstrate while increasing the number of models from $1$ to $7$ noticeably improves the accuracy, the subsequent increase from $7$ to $12$ has only marginal gains. Furthermore, the subset $\{Q_1, \dots, Q_7\}$ produces a solution with fewer artifacts than the subset $\{Q_6, \dots, Q_{12}\}$. Intuitively, as $\theta$ gets larger, $Q_{\theta}$ is closer to zero.

\begin{figure}[htbp]
    \centering
    
    \includegraphics[width=0.32\textwidth]{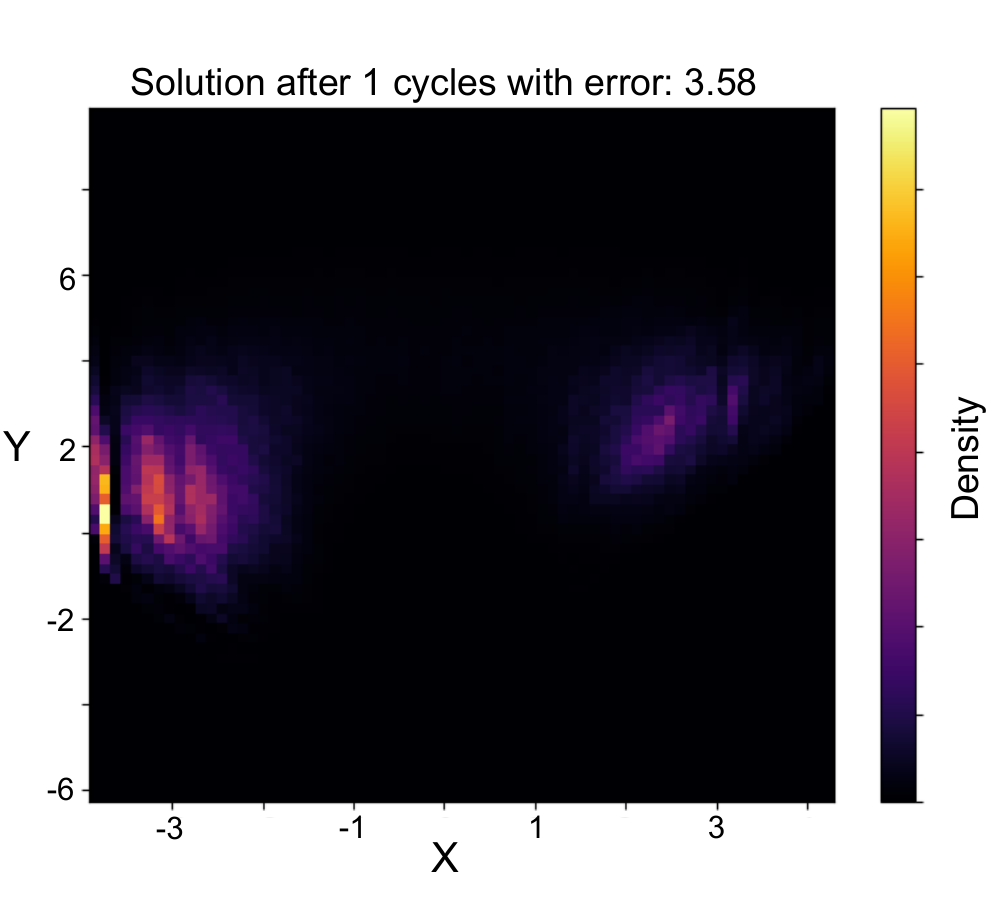}
    \hfill
    \includegraphics[width=0.32\textwidth]{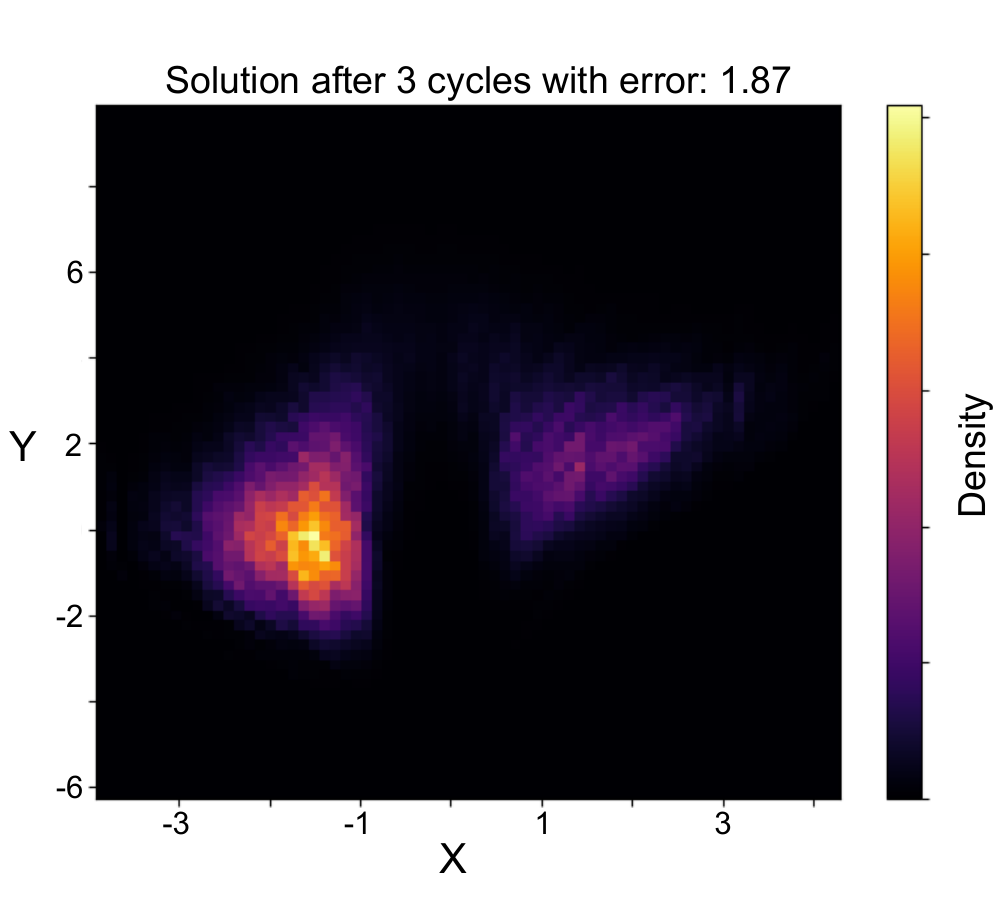}
    \hfill
    \includegraphics[width=0.32\textwidth]{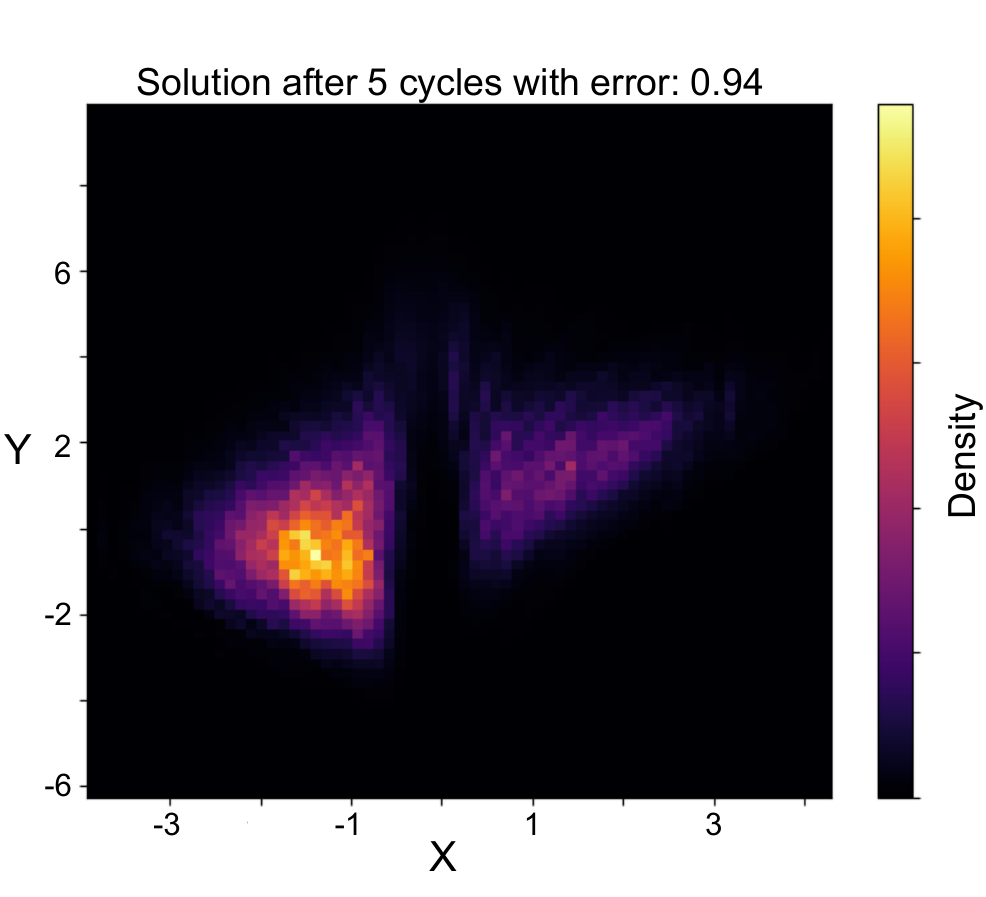}
    
    \vspace{1em} 
    
    \includegraphics[width=0.32\textwidth]{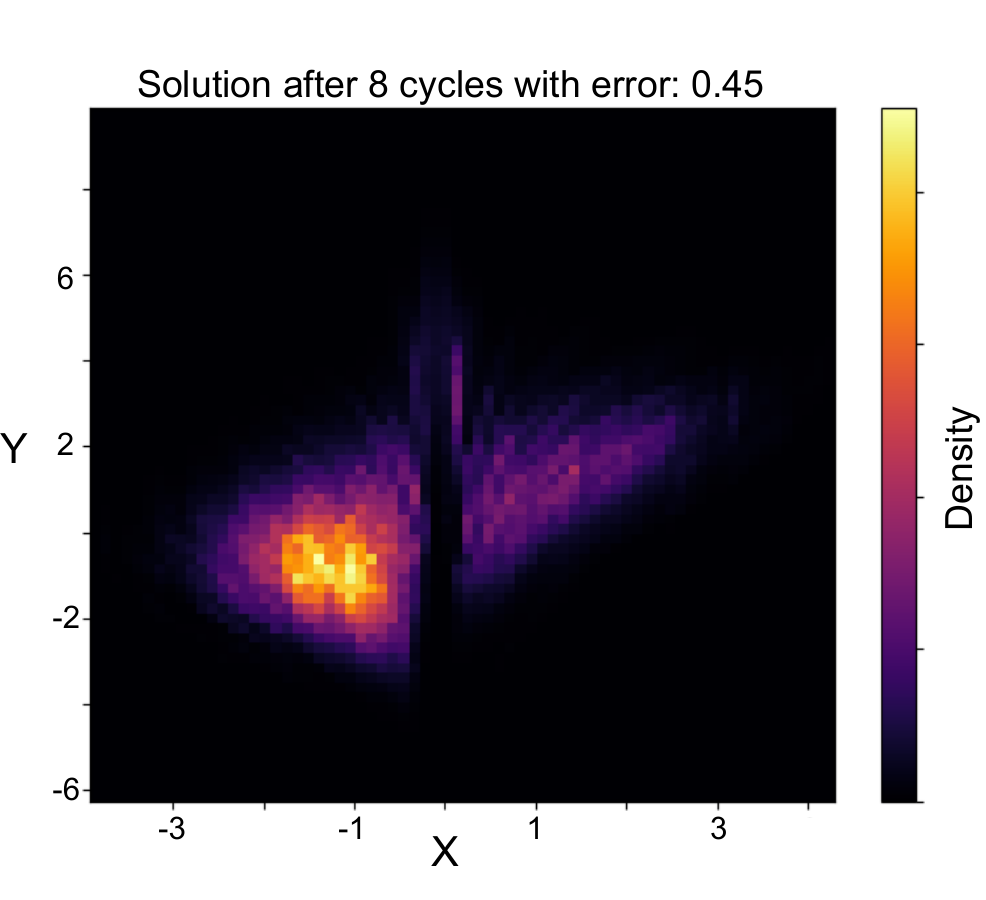}
    \hfill
    \includegraphics[width=0.32\textwidth]{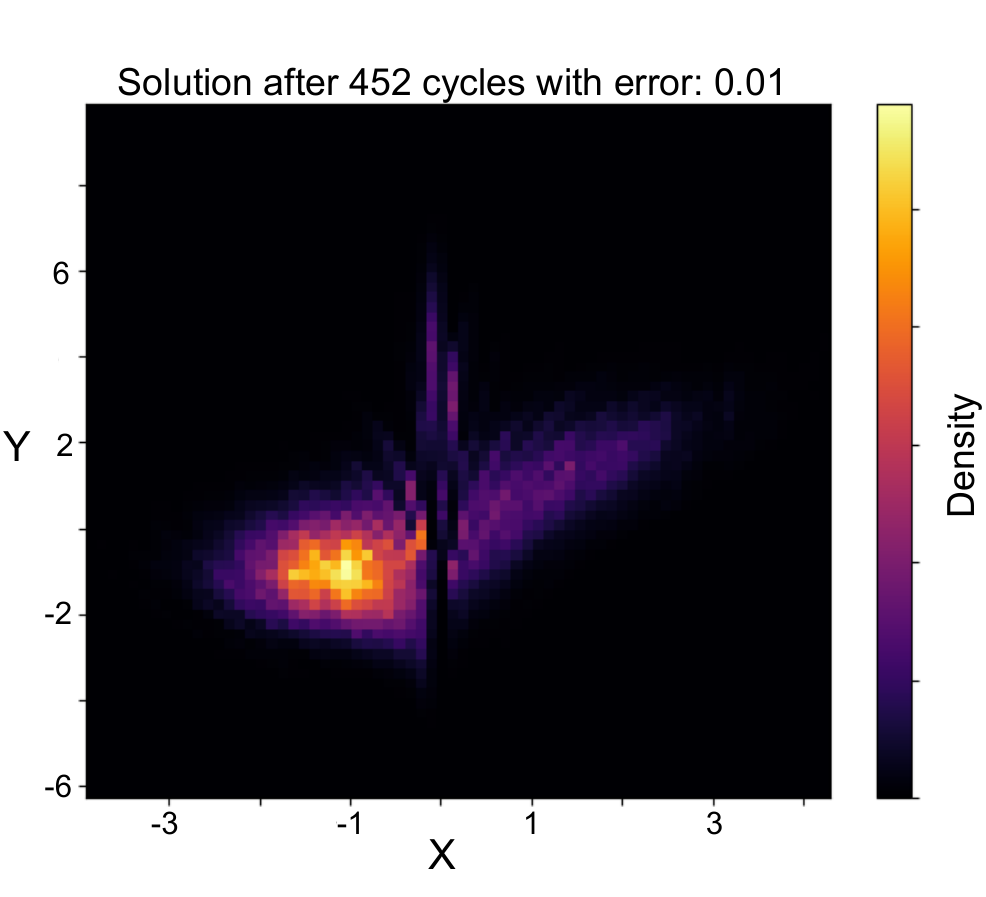}
    \hfill
    \includegraphics[width=0.32\textwidth]{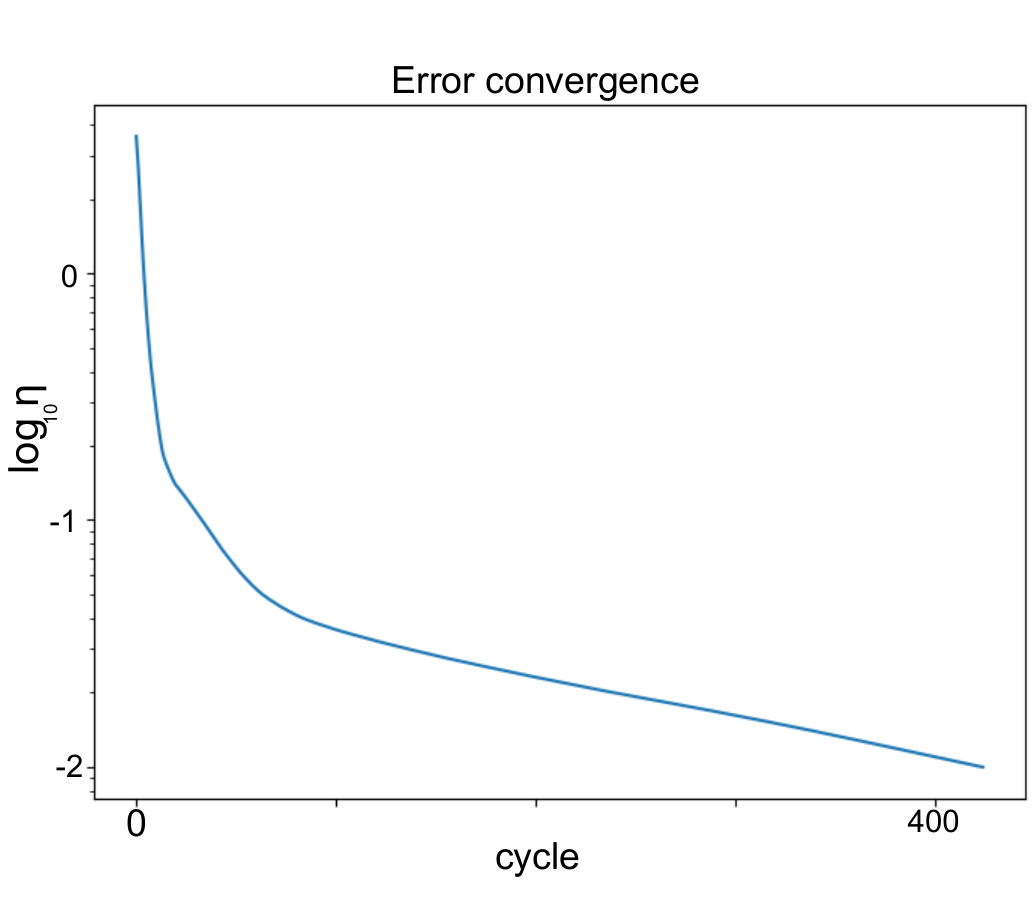}
    
    \caption{Evolution of the distribution in the second scenario at cycles $1, 3, 5, 8$, and $425$, with convergence reached at cycle $425$. The bottom-right panel shows $\log_{10}\eta$, where $\eta$ denotes the error, versus the number of cycles. }
    \label{fig: evol}
\end{figure}

In the second scenario, we assume $ P_{1,0} = \mathcal{N}\big((4,4), \big(\begin{smallmatrix}
  1 & 0\\
  0 & 1
\end{smallmatrix}\big) \big) $, which is roughly $7$ standard deviations away from the true mode $(-1,-1)$. The purpose of this scenario is to evaluate the robustness of Algorithm \ref{alg: iterative} to misspecified references. We obtain the solution using the set of models $\{Q_1, \dots Q_7\}$. Results of the solution at certain cycles are shown in Figure \ref{fig: evol}. The evolution of the solution at the first a few cycles is illustrated in Figure \ref{fig: evol}. Despite the reference distribution differing significantly from the true underlying distribution, the solution moves to the correct mode within a few cycles, showing the algorithm's robustness to inaccurate prior information. Comparing with the first scenario, the solution with the misspecified reference distribution produces more artifacts and is less accurate.

\subsection {Ensemble forecasting for Covid-19.}
We demonstrate the utility of our method through ensemble forecasting of Covid-19. Specifically, we forecast the cumulative cases of an ensemble of counties in the United States using early-stage Omicron surge data. In the following section, we briefly discuss the motivation behind our approach, along with our data and model selection. 

Since the onset of COVID-19, the flood of disease modeling has revealed systemic shortcomings in forecasting infectious diseases. Standard forecast models are often simplified, failing to account for complex issues such as shifting policies, vaccination rates, and geographic disparities. Furthermore, COVID-19 strains vary between regions, which strongly impacts both the surge rate and forecast accuracy.

To address these challenges, ensemble forecasting has become a critical tool. As a variation of the Monte Carlo method, ensemble forecasting generates a range of potential outcomes by running models multiple times sampled from a distribution of initial conditions and parameters \cite{Leutbecher,Anderson2,Gneiting, forecastflu,paireau2022ensemble,sherratt2023predictive}. Therefore, this approach might be able to characterize the variation in covid strains, for example, by learning a distribution of transmission and removal rates from data. Additionally, a set of forecasts provide a natural basis for uncertainty quantification, thereby reducing the bias associated with relying on a single, imperfect model.

\subsubsection{Data and model consideration} \label{sec: modelsetting}
\begin{figure}[htbp]    \centering
    
    \includegraphics[height=0.3\textheight]{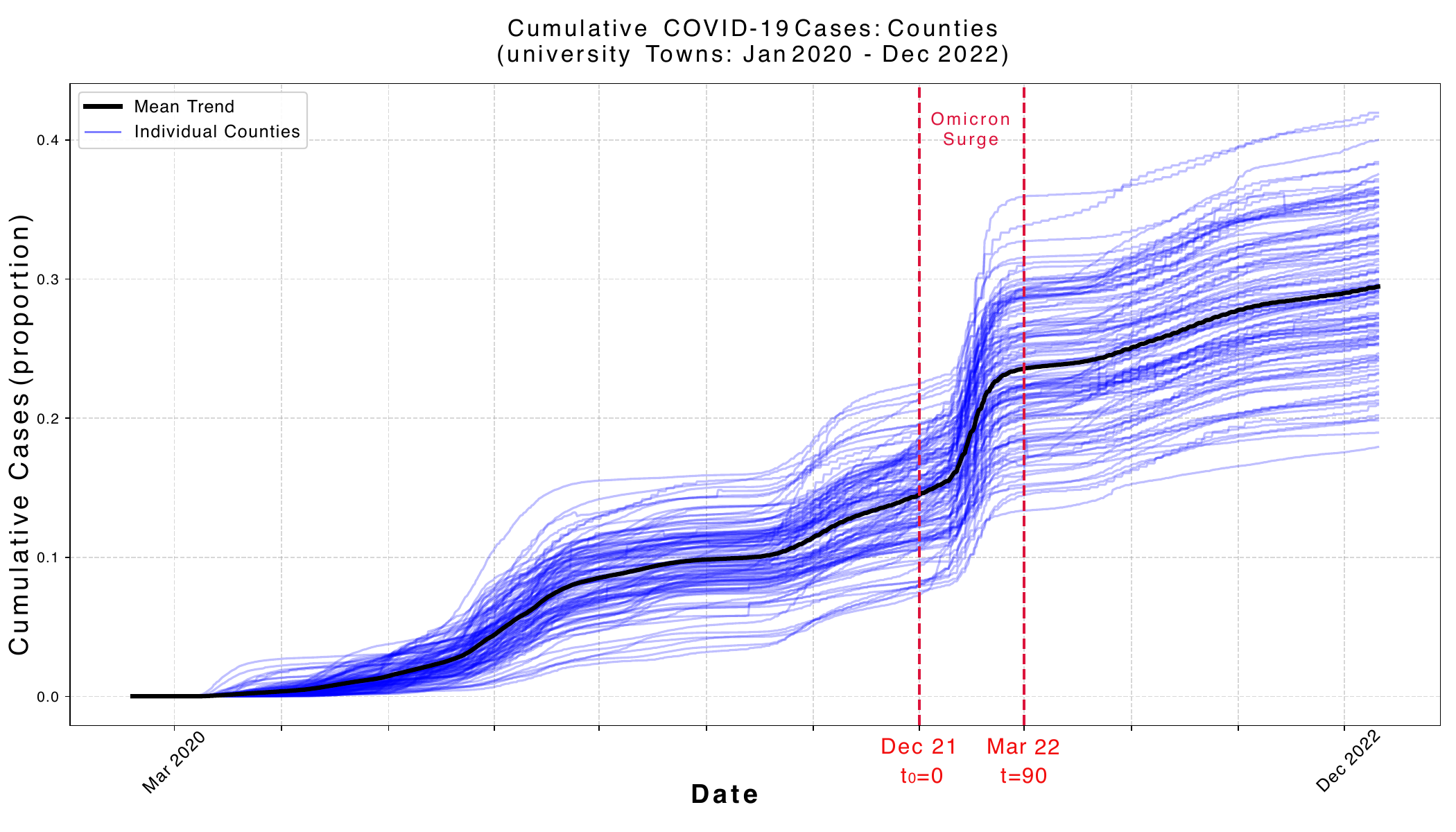}
   
       \caption{Time series of cumulative infected cases reported by The New York Times \cite{nytimes_covid19_data} from January 2020 to December 2022. The period of the Omicron surge is highlighted between the red dashed lines.}
    \label{fig: covid}
\end{figure}

 Consequently, it is crucial to sample from a carefully calibrated probability distribution of parameters and initial conditions. To achieve this, we consider a natural ensemble of different counties, utilizing their cumulative case counts at early weeks to calibrate the parameter distribution. Following the consideration discussed in \cite{kroetch2026bayesian}, a natural ensemble for analysis is 113 “university counties” in the United States that are physically isolated from large population centers. These counties share a high degree of homogeneity regarding demographics, economic structures, COVID-19 policies, and healthcare access. As seen in Figure \ref{fig: covid}, the infection trend of these university counties share similar pattern.

Consequently, our data characterizes the variability of cumulative cases across different counties at some pre-specified time points. This contrasts with longitudinal data, which tracks the trajectory of cases within a single county over time. In other words, our ``cross-sectional'' data evaluates the case variability spatially. Thus, our method differs from standard longitudinal calibration approaches such as 4D-VAR, data assimilation, Bayesian inverse problems, ensemble Kalman filtering, or particle filtering.

As for model selection, we adopt the standard SIR compartmental model, defined by the following system of ordinary differential equations:

\begin{equation} \label{eq: sir}
\frac{dS}{dt} = - \beta IS ,\,\, \frac{dI}{dt} = \beta IS - \gamma I ,\, \, \frac{dR}{dt} = \gamma I,\, t \geq t_0,
\end{equation}

with initial conditions at time $t_0$ given by $S(t_0)$, $I(t_0)$, and $R(t_0) = 1 - S(t_0) - I(t_0)$. The variables $S, I$, and $R$ represent the susceptible, infected, and removed proportions of the population, respectively, while $\beta$ and $\gamma$ denote the transmission and removal rates. The SIR model is a simple yet useful model for forecasting a single disease surge \cite{roda2020difficult,zhang2022usage}. The selected university towns possess sufficiently large and closed populations to satisfy the model’s key assumptions.

We set the starting time of the Omicron surge, $t_0 = 0$, as December 1, 2021. Since the cumulative case counts in \cite{nytimes_covid19_data} are frequently reported weekly rather than daily, we compute the cumulative case changes every 7 day, from day 7 to day 35, for each county, scaled by its total population. This yields five datasets, $\{Y_j\}_{j=1}^5$, each representing the variability in the surge rate among the counties for that specific week. Specifically, each data set $\{Y_j\}$ consists of 
$$ Y^{i}_j = \frac{\textbf{Cases}( t_0 + 7j;  \text{town i }  ) - \textbf{Cases}(t_0+7(j-1);  \text{town i }   )}{\textbf{population}\text{(town i) }}, \,\, i =1,\dots, 113.$$
Because the initial conditions $(S(t_0)$, $I(t_0)$, $R(t_0))$ and parameters $\beta, \gamma$ are unknown and vary by county, we model them as random variables. Since the Omicron surge is in the middle of pandemic following multiple infection waves and vaccinations, $S(t_0)$ may have a wide range of uncertainty. In contrast, the uncertainty of $I(t_0)$ should be small at the start of surge. We also assume a wide range value for $\beta$ and $\gamma$ to accommodate all kinds of surge dynamics.  As a result, we assume the initial distribution to be uniform: $$\big(S(t_0), I(t_0), \beta, \gamma\big) \sim \mathcal{U}\big([0.3, 0.998] \times [0, 0.002] \times [0, 3] \times [0, 3]\big).$$
Since cumulative cases equals the sum of all currently infected and removed individuals, each dataset $Y_j$ is modeled by the map $Q_j$: $$Q_j \Big(S(t_0), I(t_0), \beta, \gamma \Big) =  S(t_0 + 7(j-1)) - S\big(t_0 + 7j; S(t_0), I(t_0), \beta, \gamma\big),$$ for which $S(\cdot)$ can be numerically solved from \eqref{eq: sir}.
\begin{figure}[htbp]    \centering
    
    \includegraphics[width=0.32\textwidth]{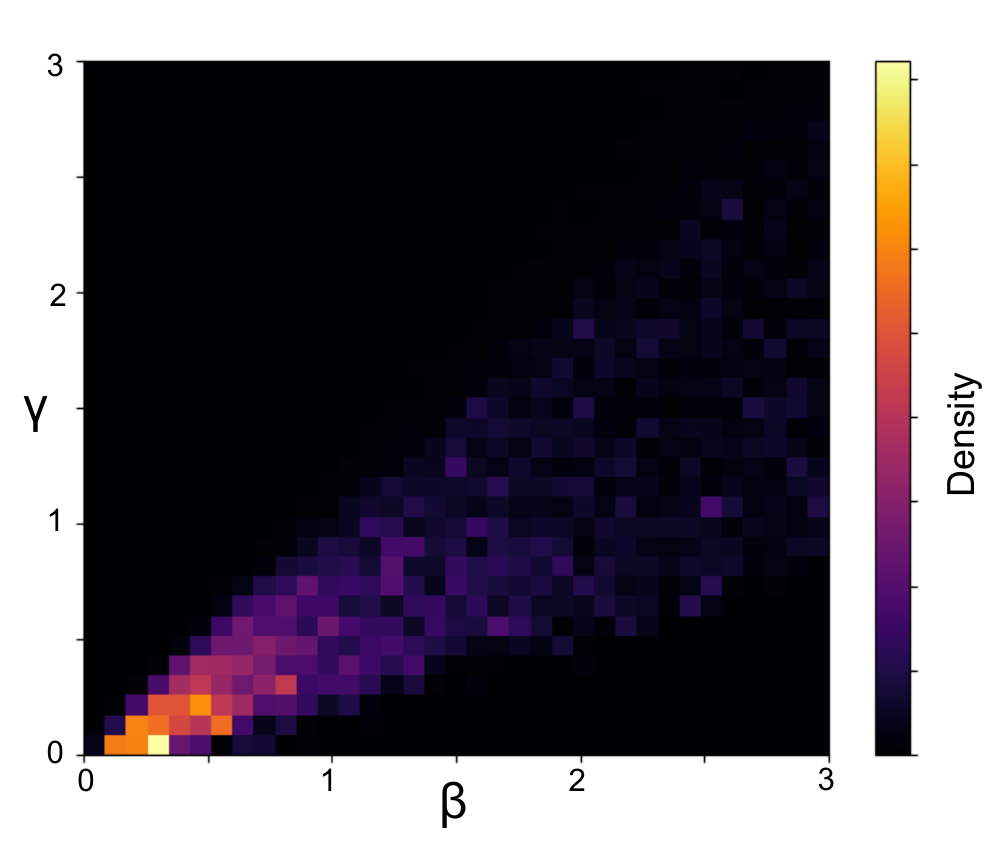}
    \hfill
    \includegraphics[width=0.32\textwidth]{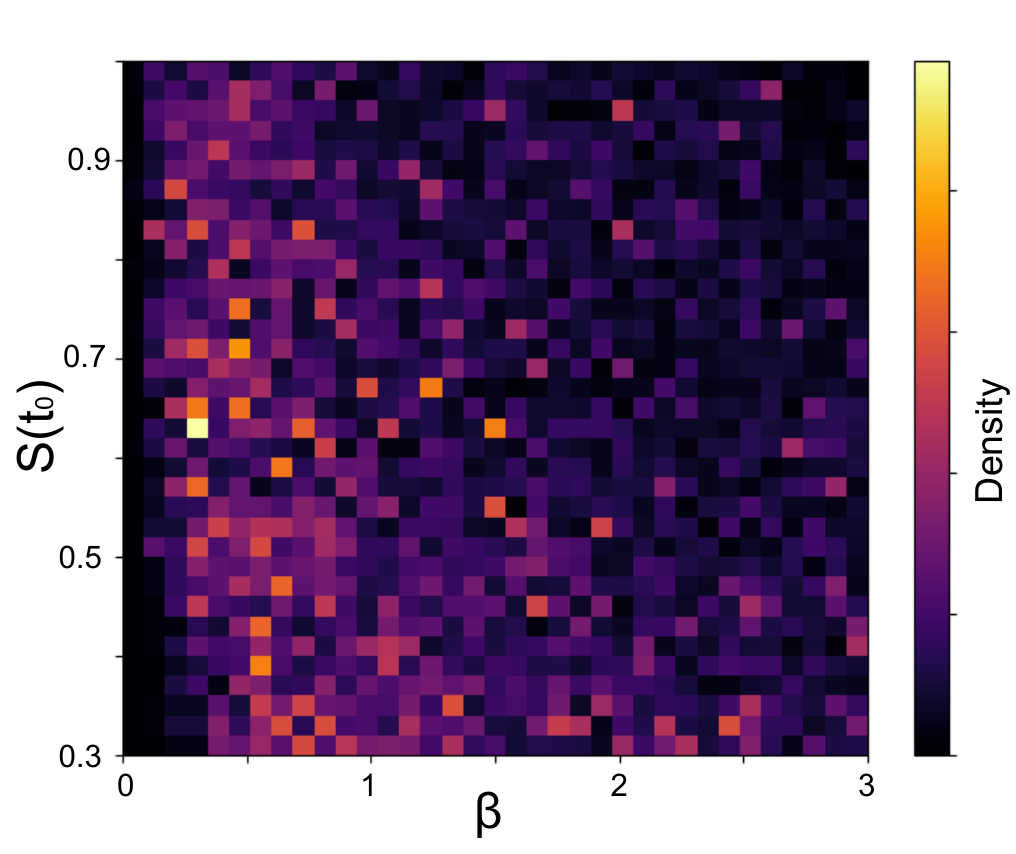}
    \hfill
    \includegraphics[width=0.32\textwidth]{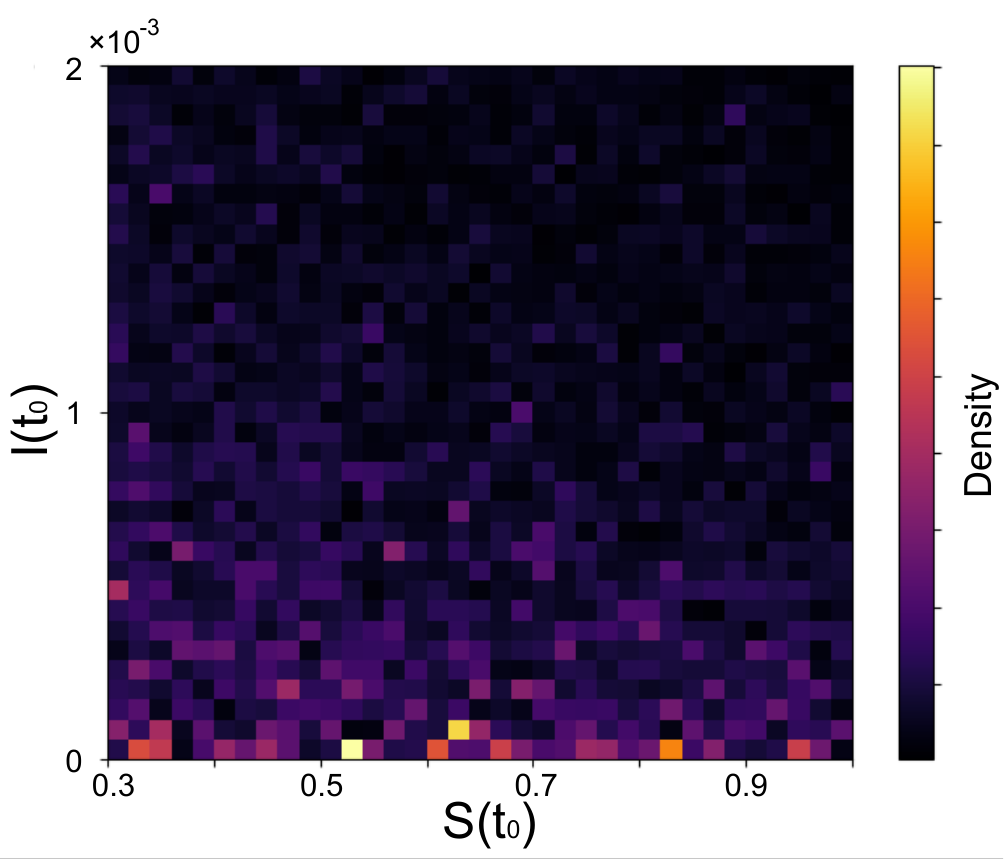}
        
    \caption{Heatmaps of the marginal distribution after convergence at $79$ cycles, with error $0.001$, using cumulative case data at days $7, 14, 21, 28$, and $35$. From left to right: the joint marginals of $(\beta,\gamma)$, $(\beta,S(t_0))$, and $(S(t_0),I(t_0))$. }
    \label{fig: marginsir}
\end{figure}
\subsubsection{Calibration and ensemble forecasting}
We  apply Algorithm \ref{alg: iterative} to calibrate a posterior probability distribution for these variables over the five datasets. Figure \ref{fig: marginsir} presents selected two-dimensional marginal distributions of the resulting posterior samples. In compartmental models, non-identifiability limits precise parameter estimation to a linear combination of $\beta$ and $\gamma$ \cite{harris2024tight,roda2020difficult,kroetch2026bayesian}. Although this linear pattern is reflected in the marginal distributions of $\beta$ and $\gamma$ (left panel of Figure \ref{fig: marginsir}), iteratively reweighting the distributions modifies the non-identifiability issue, which downweight the large value of $\beta$ and $\gamma$.

\begin{figure}[htbp]    \centering
        \includegraphics[width=0.99\textwidth]{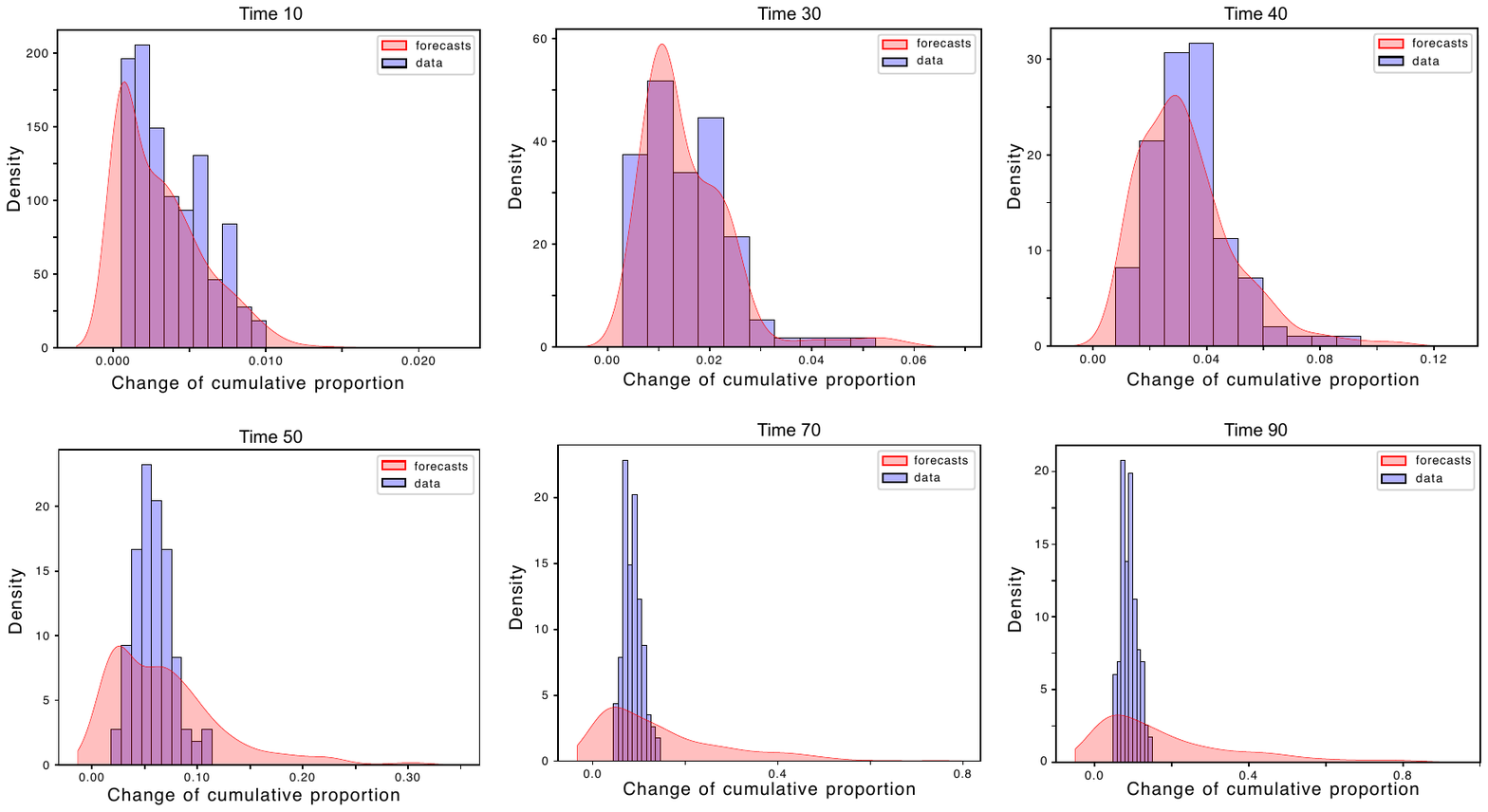}
    \caption{Comparison of the probabilistic forecast generated using the first $35$ days of data with the observed data distributions at days $10, 30, 40, 50, 70$, and $90$.}
    \label{fig: probforecast}
\end{figure}

Using the calibrated parameter distribution, we generated a probabilistic forecast for the entire duration of the surge. Figure \ref{fig: probforecast} compares the generated probability distribution against the data distribution at six different time points during the surge. Because the model was calibrated using data from the first 35 days, the probabilistic forecasts closely align with the data distribution at early time points (days 10 and 30). However, during the later stages of the surge, the true data distribution concentrates around $0.1$, whereas the forecast distribution exhibits a wider spread.  

\begin{figure}[htbp]    \centering
    
    \includegraphics[width=0.48\textwidth]{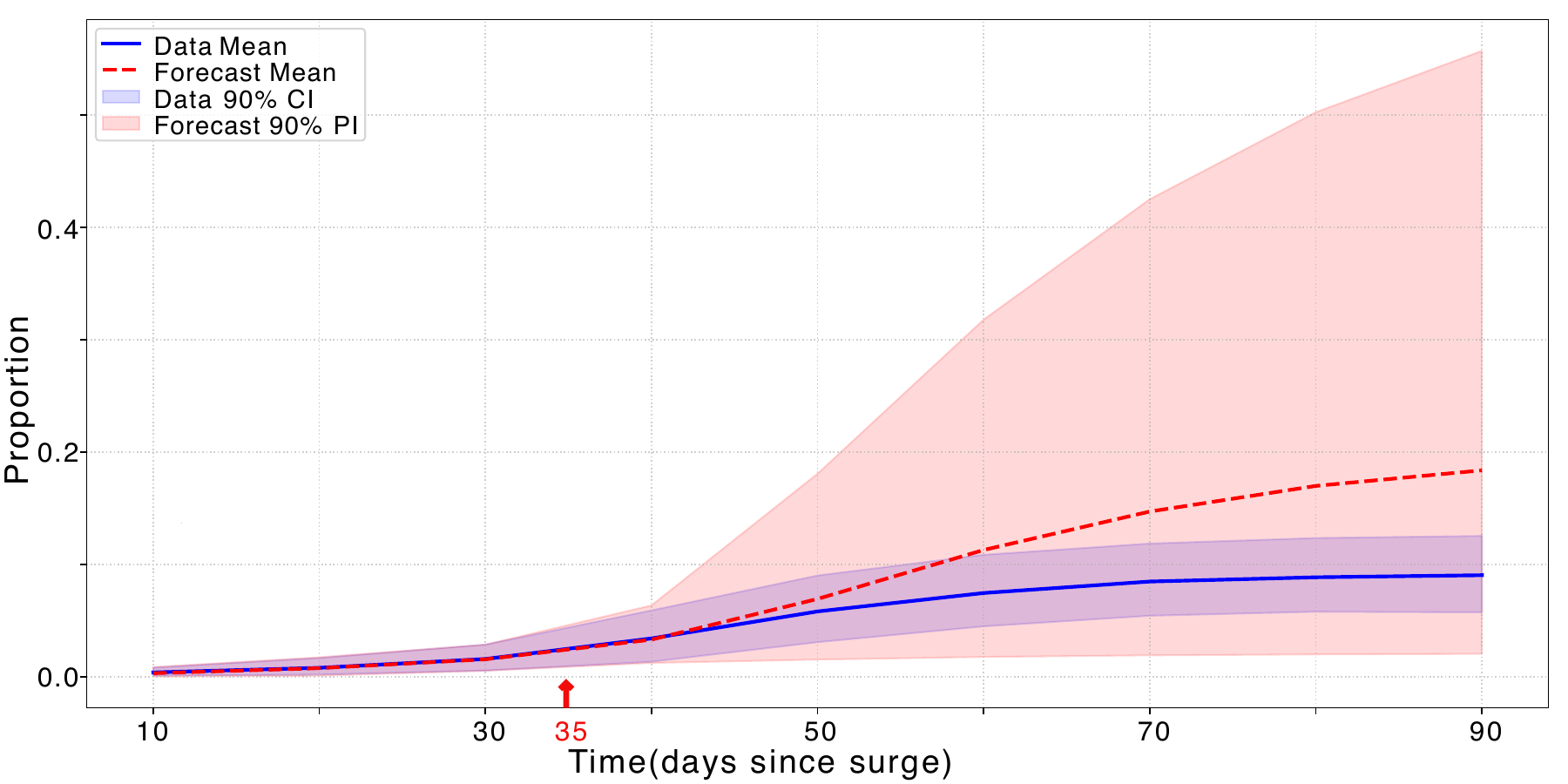}
            \hfill
      \includegraphics[width=0.48\textwidth]{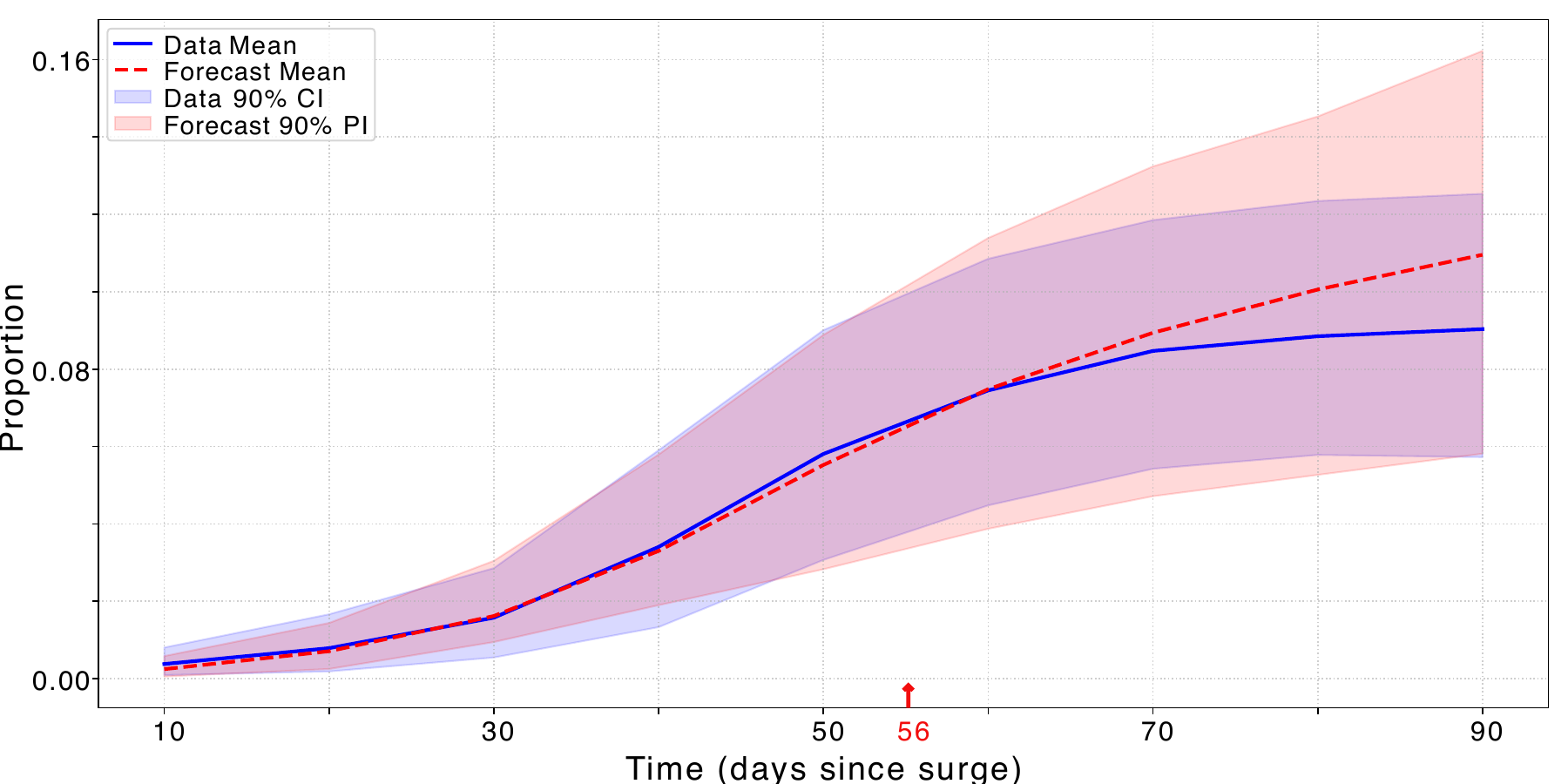}

    \caption{ Left: Comparison between the ensemble forecast (cumulative proportion increment) using \textbf{35} days data and the observed data over time. The red and blue shaded regions denote the $90\%$ quantile ranges for the forecast and the observed data, respectively. The red dashed curve and the blue curve represent the mean of the ensemble forecast and the data mean, respectively. Right: Comparison between the ensemble forecast using \textbf{56} days data and the observed data over time.  }
    \label{fig: forecastmean}
\end{figure}

As is standard practice with ensemble methods, we also evaluated the mean response of the forecast. The left panel of Figure \ref{fig: forecastmean} demonstrates that the mean response of the ensemble forecast closely aligns with the mean infection trend of the university counties. Furthermore, the uncertainty quantification band matches the actual data variability well for the first 40 days. However, the uncertainty band becomes broader than the observed data variability in the later stages. This is a reasonable forecast since the observation is limited to a very early stage (the first 35 days), which do not provide much information on how surge dynamics will evolve in the future. Thus, the solution encompasses a wide range of possibility of surge dynamics, ranging from mild scenarios where the infected proportion remains near zero, to severe outbreaks where the infected proportion reaches 0.6.

\begin{figure}[htbp]    \centering   
    \includegraphics[width=0.99\textwidth]{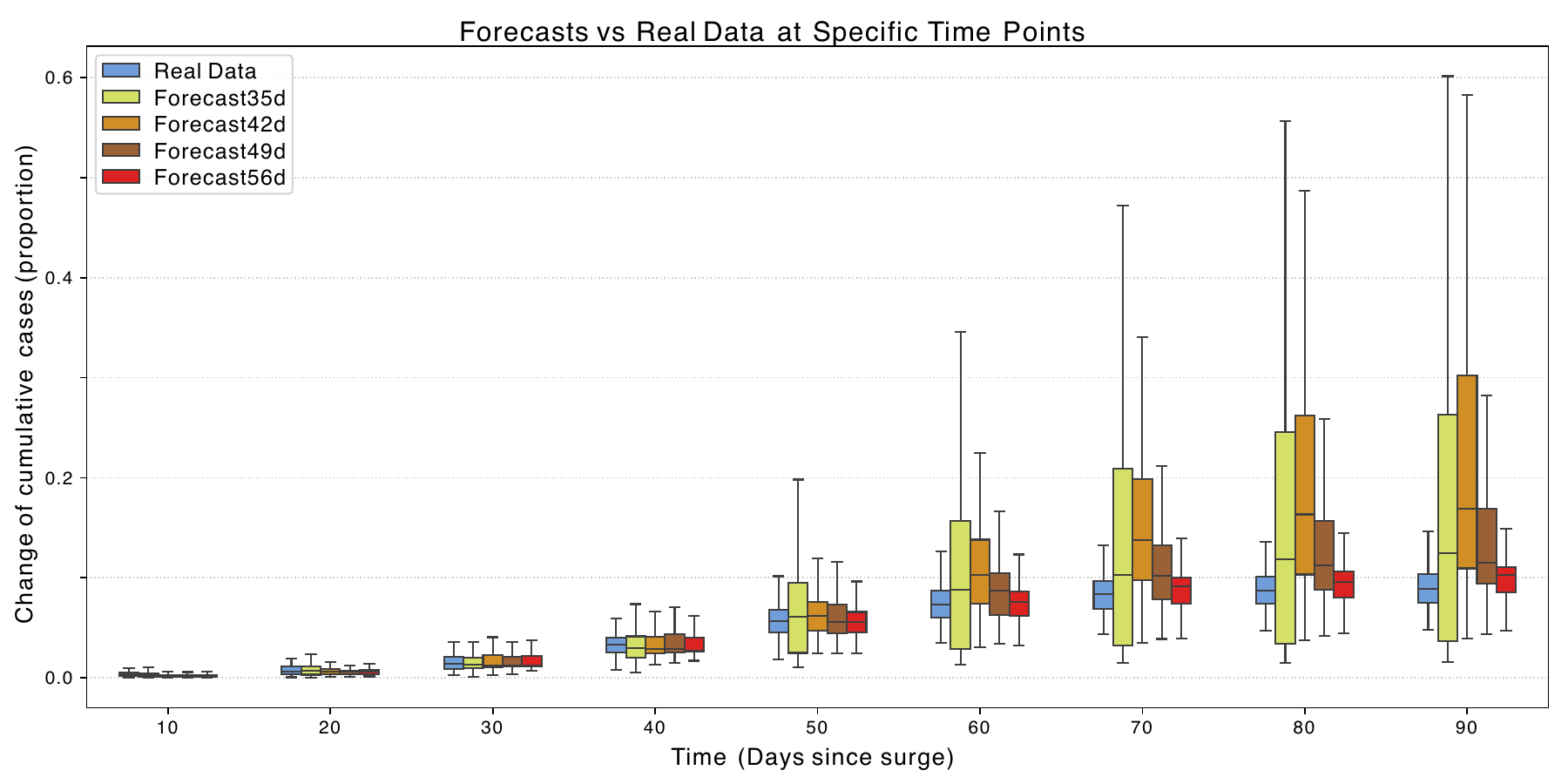}
          \caption{Boxplot comparison of the observed data and four ensemble forecasts. The Forecast35d boxplot represents the ensemble forecast using up to 35 days of data, as detailed in Section \ref{sec: modelsetting}. The Forecast42d, Forecast49d, and Forecast56d boxplots denote updated forecasts that uses up to 42, 49, and 56 days of data. }
    \label{fig: forecastbox}
\end{figure}

 To further verify this, we keep the original model settings unchanged and sequentially add three additional weeks of data (day 42, 49, 56) to compute the updated forecasts. The right panel of Figure \ref{fig: forecastmean}  compares the real data against the ensemble forecast using up to 56 days, showing a significant improvement in prediction accuracy. In addition, a comparison of the four different forecasts in Figure \ref{fig: forecastbox} indicates that the forecast's accuracy is consistent with the amount of information available on surge dynamics over time.  
 
\section{Conclusions and future directions} \label{sec: con}
This paper studies the Iterative proportional fitting procedure when the constraint sets are given by pushforward constraints. The procedure defines an iterative I-projection, which is written as a product of reference distribution $P_0$ and reweighting factors $w_{i,j}$. We establish the convergence of the sequence under regularity conditions on $P_0$ and $w_{i,j}$. To avoid the assumption on the whole sequence, we also provide a sufficient condition on $P_0$ and $\{Q\}_{j=1}^t$. The solution of IPFP is approximated by reweighing the samples from $P_0$ by $w_{i,j}$ until its convergence. The synthetic example of radioactive decay validates the algorithm's convergence and its robustness to misspecified $P_0$, and also implies the solution depends on the number of and the arrangement of  $\{Q_j\}_{j=1}^{t}$. Finally, we conduct ensemble forecasting for Omicron surge. By considering 113 university towns as an ensemble, we calibrate a probability distribution on parameters of SIR model using earlier stage data and use the distribution to forecast later stage. The distribution on transmission and removed rate may provide insight on the variation of covid strains. Accounting for such variation is important for forecasting the surge dynamics. 

For future directions,
the sequential importance sampling algorithm can be computationally inefficient when the dimensionality of model parameter is very high. We are interested in tackling this difficulty in future work. In addition, the synthetic example shows that different combinations of $\{Q\}_j$ could impact the convergence rate and solution accuracy of the algorithm. This could lead to experimental design problem for finding an optimal arrangement that yields faster convergence and more accurate solutions. Finally, since this multi-pushforward constraints problem has a multi-marginal Shr\"odinger bridge interpretation, we would like to explore how this IPFP construction could be helpful for constructing stochastic processes on the path space. The model calibration examples in this paper can be thought of constructing a stochastic process induced by random parameter in $\mathbb{R}^l$. Whether or not the iterative construction is applicable for general stochastic processes needs further investigation.  

\section{Acknowledgement}
H. Shi acknowledges the support of the Natural Sciences and Engineering Research Council of Canada.
D. Estep acknowledges the support of the Canada Research
Chairs Program, the Natural Sciences and Engineering Research Council of Canada, Canada’s Digital Technology
Superclustre, the Dynamics Research Corporation, the Idaho National Laboratory, the U.S. Department of Energy, the
U.S. National Institutes of Health, the U.S. National Science Foundation, and Riverside Research. 
\section{Data availability}
The code for reproducing the radioactive decay simulation and the COVID-19 ensemble forecasting application is publicly available at \url{https://github.com/HaiyiS/IPFP_SCP}. The cumulative COVID-19 case data are provided by The New York Times and are available at \url{https://github.com/nytimes/covid-19-data}.

\bibliographystyle{abbrv}
\bibliography{refs_iterative}

\begin{thebibliography}{10}

\bibitem{anden2018structural}
J.~And{\'e}n and A.~Singer.
\newblock Structural variability from noisy tomographic projections.
\newblock {\em SIAM Journal on Imaging Sciences}, 11(2):1441--1492, 2018.

\bibitem{Anderson2}
J.~Anderson and S.~Anderson.
\newblock A {Monte Carlo} implementation of the nonlinear filtering problem to
  produce ensemble assimilations and forecasts.
\newblock {\em Monthly Weather Review}, 127:2741--2758, 1999.

\bibitem{Estepprob}
S.~Basu, T.~Butler, D.~Estep, and N.~Panda.
\newblock {\em A ramble through probability: How {I} learned to stop worrying
  and love measure theory}.
\newblock Society for Industrial and Applied Mathematics, Philadelphia, PA,
  2024.

\bibitem{baumhofer2014production}
T.~Baumh{\"o}fer, M.~Br{\"u}hl, S.~Rothgang, and D.~U. Sauer.
\newblock Production caused variation in capacity aging trend and correlation
  to initial cell performance.
\newblock {\em Journal of Power Sources}, 247:332--338, 2014.

\bibitem{bhattacharya2006iterative}
B.~Bhattacharya.
\newblock An iterative procedure for general probability measures to obtain
  {I}-projections onto intersections of convex sets.
\newblock {\em The Annals of Statistics}, 2006.

\bibitem{bingham2024inverse}
D.~Bingham, T.~Butler, and D.~Estep.
\newblock Inverse problems for physics-based process models.
\newblock {\em Annual Review of Statistics and Its Application}, 11, 2024.

\bibitem{bishop1969incomplete}
Y.~M.~M. Bishop and S.~E. Fienberg.
\newblock Incomplete two-dimensional contingency tables.
\newblock {\em Biometrics}, pages 119--128, 1969.

\bibitem{Bog07}
V.~I. Bogachev.
\newblock {\em Measure theory}, volume~1.
\newblock Springer-Verlag, Berlin, 2007.

\bibitem{breidt2011measure}
J.~Breidt, T.~Butler, and D.~Estep.
\newblock A measure-theoretic computational method for inverse sensitivity
  problems {I}: Method and analysis.
\newblock {\em SIAM Journal on Numerical Analysis}, 49(5):1836--1859, 2011.

\bibitem{britton2013experimentally}
O.~J. Britton, A.~Bueno-Orovio, K.~Van~Ammel, H.~R. Lu, R.~Towart, D.~J.
  Gallacher, and B.~Rodriguez.
\newblock Experimentally calibrated population of models predicts and explains
  intersubject variability in cardiac cellular electrophysiology.
\newblock {\em Proceedings of the National Academy of Sciences},
  110(23):E2098--E2105, 2013.

\bibitem{butler2012computational}
T.~Butler, D.~Estep, and J.~Sandelin.
\newblock A computational measure theoretic approach to inverse sensitivity
  problems {II}: A posteriori error analysis.
\newblock {\em SIAM Journal on Numerical Analysis}, 50(1):22--45, 2012.

\bibitem{butler2014measure}
T.~Butler, D.~Estep, S.~Tavener, C.~Dawson, and J.~J. Westerink.
\newblock A measure-theoretic computational method for inverse sensitivity
  problems {III}: Multiple quantities of interest.
\newblock {\em SIAM/ASA Journal on Uncertainty Quantification}, 2(1):174--202,
  2014.

\bibitem{Butler_2020}
T.~Butler, T.~Wildey, and T.~Y. Yen.
\newblock Data-consistent inversion for stochastic input-to-output maps.
\newblock {\em Inverse Problems}, 36(8):085015, August 2020.

\bibitem{chi2021sliced}
J.~Chi.
\newblock {\em Sliced inverse approach and domain recovery for stochastic
  inverse problems}.
\newblock PhD thesis, Colorado State University, 2021.

\bibitem{csiszar1975divergence}
I.~Csisz{\'a}r.
\newblock {I}-divergence geometry of probability distributions and minimization
  problems.
\newblock {\em The Annals of Probability}, pages 146--158, 1975.

\bibitem{cuturi2013sinkhorn}
M.~Cuturi.
\newblock {Sinkhorn} distances: Lightspeed computation of optimal transport.
\newblock In {\em Advances in Neural Information Processing Systems},
  volume~26, 2013.

\bibitem{david2004order}
H.~A. David and H.~N. Nagaraja.
\newblock {\em Order statistics}.
\newblock John Wiley \& Sons, 2004.

\bibitem{de2021diffusion}
V.~De~Bortoli, J.~Thornton, J.~Heng, and A.~Doucet.
\newblock Diffusion schr$\backslash$" odinger bridge with applications to
  score-based generative modeling.
\newblock {\em arXiv preprint arXiv:2106.01357}, 2021.

\bibitem{dechent2021estimation}
P.~Dechent, S.~Greenbank, F.~Hildenbrand, S.~Jbabdi, D.~U. Sauer, and D.~A.
  Howey.
\newblock Estimation of {Li-ion} degradation test sample sizes required to
  understand cell-to-cell variability.
\newblock {\em Batteries \& Supercaps}, 4(12):1821--1829, 2021.

\bibitem{deming1940least}
W.~E. Deming and F.~F. Stephan.
\newblock On a least squares adjustment of a sampled frequency table when the
  expected marginal totals are known.
\newblock {\em The Annals of Mathematical Statistics}, 11(4):427--444, 1940.

\bibitem{tol2001}
A.~Di~Bucchianico, J.~H. Einmahl, and N.~A. Mushkudiani.
\newblock Smallest nonparametric tolerance regions.
\newblock {\em The Annals of Statistics}, 29(5):1320--1343, 2001.

\bibitem{dimarino2020optimal}
S.~Di~Marino and A.~Gerolin.
\newblock An optimal transport approach for the {Schr{\"o}dinger} bridge
  problem and convergence of {Sinkhorn} algorithm.
\newblock {\em Journal of Scientific Computing}, 85(2):27, 2020.

\bibitem{dykstra1985iterative}
R.~L. Dykstra.
\newblock An iterative procedure for obtaining {I}-projections onto the
  intersection of convex sets.
\newblock {\em The Annals of Probability}, pages 975--984, 1985.

\bibitem{espinosa2025cryo}
D.~S. Espinosa, E.~H. Thiede, and Y.~Yang.
\newblock {Cryo-EM} as a stochastic inverse problem.
\newblock arXiv preprint, arXiv:2509.05541, 2025.

\bibitem{fienberg1970iterative}
S.~E. Fienberg.
\newblock An iterative procedure for estimation in contingency tables.
\newblock {\em The Annals of Mathematical Statistics}, 41(3):907--917, 1970.

\bibitem{Gneiting}
T.~Gneiting, F.~Balabdaoui, and A.~Raftery.
\newblock Probabilistic forecasts, calibration and sharpness.
\newblock {\em Journal of the Royal Statistical Society: Series B (Statistical
  Methodology)}, 69:243--268, 2007.

\bibitem{harris2024tight}
J.~E. Harris.
\newblock A tight fit of the {SIR} dynamic epidemic model to daily cases of
  {COVID-19} reported during the 2021--2022 omicron surge in {New York City}: A
  novel approach.
\newblock {\em Statistical Methods in Medical Research}, 33(10):1877--1898,
  2024.

\bibitem{harris2017failure}
S.~J. Harris, D.~J. Harris, and C.~Li.
\newblock Failure statistics for commercial lithium ion batteries: A study of
  24 pouch cells.
\newblock {\em Journal of Power Sources}, 342:589--597, 2017.

\bibitem{jiang2026iterative}
T.~Jiang, T.~Butler, T.~Wildey, T.~Kutta, and H.~Wang.
\newblock Iterative data-consistent inversion with multiple push-forward
  constraints.
\newblock {\em International Journal for Uncertainty Quantification}, 16(3),
  2026.

\bibitem{kroetch2026bayesian}
K.~Kroetch and D.~Estep.
\newblock {Bayesian} inverse ensemble forecasting for {COVID-19}.
\newblock {\em Canadian Journal of Statistics}, 2026.

\bibitem{kullback1968probability}
S.~Kullback.
\newblock Probability densities with given marginals.
\newblock {\em The Annals of Mathematical Statistics}, 39(4):1236--1243, 1968.

\bibitem{lahr2004biproportional}
M.~Lahr and L.~De~Mesnard.
\newblock Biproportional techniques in input-output analysis: Table updating
  and structural analysis.
\newblock {\em Economic Systems Research}, 16(2):115--134, 2004.

\bibitem{lawson2018unlocking}
B.~A.~J. Lawson, C.~C. Drovandi, N.~Cusimano, P.~Burrage, B.~Rodriguez, and
  K.~Burrage.
\newblock Unlocking data sets by calibrating populations of models to data
  density: A study in atrial electrophysiology.
\newblock {\em Science Advances}, 4(1):e1701676, 2018.

\bibitem{leonard2014survey}
C.~L{\'e}onard.
\newblock A survey of the {Schr{\"o}dinger} problem and some of its connections
  with optimal transport.
\newblock {\em Discrete and Continuous Dynamical Systems}, 34(4):1533--1574,
  2014.

\bibitem{Leutbecher}
M.~Leutbecher and T.~N. Palmer.
\newblock Ensemble forecasting.
\newblock {\em Journal of Computational Physics}, 227:3515--3539, 2008.

\bibitem{li2025inverse}
Q.~Li, M.~Oprea, L.~Wang, and Y.~Yang.
\newblock Inverse problems over probability measure space.
\newblock arXiv preprint, arXiv:2504.18999, 2025.

\bibitem{Li2024inverse}
Q.~Li, L.~Wang, and Y.~Yang.
\newblock Differential equation--constrained optimization with stochasticity.
\newblock {\em SIAM/ASA Journal on Uncertainty Quantification}, 12(2):549--578,
  2024.

\bibitem{marder2011multiple}
E.~Marder and A.~L. Taylor.
\newblock Multiple models to capture the variability in biological neurons and
  networks.
\newblock {\em Nature Neuroscience}, 14(2):133--138, 2011.

\bibitem{Mosegaard1996}
K.~Mosegaard and A.~Tarantola.
\newblock {Monte Carlo} sampling of solutions to inverse problems.
\newblock {\em Journal of Geophysical Research: Solid Earth},
  100(B7):12431--12447, 1995.

\bibitem{Mosegard2002}
K.~Mosegaard and A.~Tarantola.
\newblock Probabilistic approach to inverse problems.
\newblock In W.~Lee, H.~Kanamori, P.~Jennings, and C.~Kisslinger, editors, {\em
  International Handbook of Earthquake and Engineering Seismology, Part {A}},
  volume~81 of {\em International Geophysics}, pages 237--265. Academic Press,
  2002.

\bibitem{paireau2022ensemble}
J.~Paireau, A.~Andronico, N.~Hoz{\'e}, M.~Layan, P.~Crepey, A.~Roumagnac,
  M.~Lavielle, P.-Y. Bo{\"e}lle, and S.~Cauchemez.
\newblock An ensemble model based on early predictors to forecast {COVID-19}
  health care demand in {France}.
\newblock {\em Proceedings of the National Academy of Sciences},
  119(18):e2103302119, 2022.

\bibitem{penczek2006estimation}
P.~A. Penczek, C.~Yang, J.~Frank, and C.~M.~T. Spahn.
\newblock Estimation of variance in single-particle reconstruction using the
  bootstrap technique.
\newblock {\em Journal of Structural Biology}, 154(2):168--183, 2006.

\bibitem{PR2000}
D.~Poole and A.~Raftery.
\newblock Inference for deterministic simulation models: The {Bayesian} melding
  approach.
\newblock {\em Journal of the American Statistical Association}, 95:1244--1255,
  2000.

\bibitem{prasadan2026non}
A.~Prasadan, S.~Basu, F.~Yazdi, D.~Bingham, and D.~Estep.
\newblock Non-parametric model calibration with stochastic control parameters.
\newblock arXiv preprint, arXiv:2607.16975, 2026.

\bibitem{roda2020difficult}
W.~C. Roda, M.~B. Varughese, D.~Han, and M.~Y. Li.
\newblock Why is it difficult to accurately predict the {COVID-19} epidemic?
\newblock {\em Infectious Disease Modelling}, 5:271--281, 2020.

\bibitem{ruschendorf1995convergence}
L.~R{\"u}schendorf.
\newblock Convergence of the iterative proportional fitting procedure.
\newblock {\em The Annals of Statistics}, pages 1160--1174, 1995.

\bibitem{forecastflu}
J.~Shaman and A.~Karspeck.
\newblock Forecasting seasonal outbreaks of influenza.
\newblock {\em Proceedings of the National Academy of Sciences},
  109(50):20425--20430, 2012.

\bibitem{sherratt2023predictive}
K.~Sherratt, H.~Gruson, R.~Grah, H.~Johnson, R.~Niehus, B.~Prasse, F.~Sandmann,
  J.~Deuschel, D.~Wolffram, S.~Abbott, et~al.
\newblock Predictive performance of multi-model ensemble forecasts of
  {COVID-19} across {European} nations.
\newblock {\em eLife}, 12:e81916, 2023.

\bibitem{shi2025nonparametric}
H.~Shi, L.~Yang, J.~Chi, T.~Butler, H.~Wang, D.~Bingham, and D.~Estep.
\newblock Nonparametric {Bayesian} calibration of computer models.
\newblock arXiv preprint, arXiv:2509.22597, 2025.

\bibitem{sinkhorn1967diagonal}
R.~Sinkhorn.
\newblock Diagonal equivalence to matrices with prescribed row and column sums.
\newblock {\em The American Mathematical Monthly}, 74(4):402--405, 1967.

\bibitem{pmlr-v206-stromme23a}
A.~Stromme.
\newblock Sampling from a {Schr{\"o}dinger} bridge.
\newblock In {\em Proceedings of the 26th International Conference on
  Artificial Intelligence and Statistics}, Proceedings of Machine Learning
  Research, pages 4058--4067. PMLR, 2023.

\bibitem{Stuart_Bayesian}
A.~Stuart.
\newblock Inverse problems: A {Bayesian} perspective.
\newblock {\em Acta Numerica}, 19:451--559, 2010.

\bibitem{nytimes_covid19_data}
{The New York Times}.
\newblock Coronavirus ({COVID-19}) data in the {United States}, 2021.

\bibitem{vadeboncoeur2026efficient}
A.~Vadeboncoeur, M.~Girolami, and A.~M. Stuart.
\newblock Efficient deconvolution in populational inverse problems.
\newblock {\em International Journal for Numerical Methods in Engineering},
  127(9):e70335, 2026.

\bibitem{wilkinson2023introduction}
R.~D. Wilkinson and C.~W. Lanyon.
\newblock An introduction to the calibration of computer models.
\newblock arXiv preprint, arXiv:2310.09214, 2023.

\bibitem{yang2018infinite}
L.~Yang.
\newblock {\em Infinite dimensional stochastic inverse problems}.
\newblock PhD thesis, Colorado State University, 2018.

\bibitem{young2010tolerance}
D.~S. Young.
\newblock Tolerance: An {R} package for estimating tolerance intervals.
\newblock {\em Journal of Statistical Software}, 36:1--39, 2010.

\bibitem{zhang2022usage}
P.~Zhang, K.~Feng, Y.~Gong, J.~Lee, S.~Lomonaco, and L.~Zhao.
\newblock Usage of compartmental models in predicting {COVID-19} outbreaks.
\newblock {\em The AAPS Journal}, 24(5):98, 2022.

\end{thebibliography}

\section{Proofs} \label{sec: prof}

\subsection{Proof of Lemma~\ref{lem: abscon}}
\label{proof: abscon}
\begin{proof}
We first show by induction that $\barP \ll P_{i,j}$ for all $i,j$. By Assumption \ref{assp: abscon}, this holds for the base case $P_{1,0}$. Assume $\barP \ll P_{i,j-1}$, so $\Qkp \barP \ll \Qkp P_{i,j-1}$ for every $k$. In particular, by taking $k =j$, $\Qjp \barP = \PMDj \ll \Qjp P_{i,j-1}$.  
 Let $A \subseteq \Lambda$ be any measurable set such that $P_{i,j}(A) = 0$. By definition, $$P_{i,j}(A) = \int_A \frac{d \PMDj}{d \Qjp P_{i,j-1}}(Q_j(\lambda)) \, dP_{i,j-1} = 0.$$ Because the integrand is non-negative, it is zero $P_{i,j-1}$-a.e. on $A$. Let $Z = \{y \in \MD_j : \frac{d \PMDj}{d \Qjp P_{i,j-1}}(y) = 0\}$ and then $A \subseteq Q_j^{-1}(Z) \cup N$ for some $N$ such that $P_{i,j-1}(N) = 0$. Because $\barP \ll P_{i,j-1}$, we have $\barP(N) = 0$. Furthermore, because $\barP \in \MC_j$, $$\barP(Q_j^{-1}(Z)) = \Qjp\barP(Z) = \PMDj(Z) = \int_Z \frac{d \PMDj}{d \Qjp P_{i,j-1}} \, d(\Qjp P_{i,j-1}) = 0. $$Therefore, $\barP(A) \leq \barP(Q_j^{-1}(Z)) + \barP(N) = 0$ and hence $\barP \ll P_{i,j}$. Consequently, the absolute continuity of its pushforward is also preserved: $\PMDk \ll \Qkp P_{i,j}$ for every $k$.
\end{proof}
\subsection{Proof of Corollary~\ref{cor: equivalence}}
\label{proof: equivalence}

\begin{proof}
By definition \eqref{eq: Piproj},
 $P_{i,j}\ll P_{i,j-1}$ for every $i,j$.
Thus every later iterate is absolutely continuous with respect
to every earlier iterate. In particular, for $i\geq 2$, $P_{i,j-1}\ll P_{1,j}.$
Taking pushforward by $Q_j$ and using Lemma~\ref{lem: changeov}, we have
\[
    \Qjp P_{i,j-1}
    \ll \Qjp P_{1,j}
    =\PMDj.
\]
Lemma~\ref{lem: abscon} gives the reverse absolute continuity,
and hence $\PMDj\sim\Qjp P_{i,j-1}.$
Consequently,
\[
    0<w_{i,j}
    :=\frac{d\PMDj}{d(\Qjp P_{i,j-1})}<\infty
    \qquad \Qjp P_{i,j-1}\text{-a.e.},\]
which implies $ 0<w_{i,j}(\Qjl)<\infty$ $P_{i,j-1}${-a.e.}
Therefore, for every measurable $A\subseteq\Lambda$,
\[
    P_{i,j}(A)
    =\int_A w_{i,j}(\Qjl)\,dP_{i,j-1}=0
    \quad\Longleftrightarrow\quad
    P_{i,j-1}(A)=0.
\]
This proves $P_{i,j}\sim P_{i,j-1}$ for every $i\geq 2$.

Finally, we have
\[
    P_{1,t}=P_{2,0}
    \sim P_{2,1}\sim\cdots\sim P_{2,t}
    =P_{3,0}\sim\cdots.
\]
This means $P_{i,j}$ are mutually absolute continuous for $i >1$.
\end{proof}

\subsection{Proof of Lemma~\ref{lem: anyP}}
\label{proof: anyP}

\begin{proof}
 Let the integrand be $f(\lambda) =  \ln \frac{d\PMDj}{d\Qjp P_{i,j-1}}(\Qjl)$. To show the integral is well-defined, we decompose $f(\lambda)$ into its positive and negative parts, $f = f^+ - f^-$,
and apply the change of variable formula  to each part separately. 
Integrating $f^+(\lambda)$ over $\Lambda$ yields:
\begin{eqnarray}
\IL f^+(\lambda) \, d\barP_j &=& \IL  \max\left( \ln \frac{d\PMDj}{d \Qjp P_{i,j-1}}(\Qjl), 0 \right) \, d\barP_j
  \nonumber \\
&=& \IDj \max\left( \ln \frac{d \PMDj}{d \Qjp P_{i,j-1}}(y_j), 0 \right) \, d \Qjp \barP_j \nonumber \\
&=& \IDj \max\left( \ln \frac{d \PMDj}{d \Qjp P_{i,j-1}}(y_j), 0 \right) \, d \PMDj \label{eq: pos_part},
\end{eqnarray}
where the last equality follows from $\barP_j \in \MC_j$.
Similarly, the integral of the negative part evaluates to:
\begin{equation} \label{eq: neg}
\IL f^- \, d\barP_j =\IDj \max\left(-\ln \frac{d \PMDj}{d \Qjp P_{i,j-1}}(y_j), 0 \right) \, d \PMDj
\end{equation}
Let $x = \frac{d \PMDj}{d \Qjp P_{i,j-1}}$. The integrand $\ln x$ in \eqref{eq: neg} is non-zero only when the ratio $x<1$. Substituting $d \PMDj = x  \, d \Qjp P_{i,j-1}$ and using the bound $x(-\ln x) \leq 1/e$ for $x \in (0,1)$, 
\begin{equation}
\IL f^-(\lambda) \, d\barP_j \leq \IDj  \frac{1}{e}  \, d \Qjp P_{i,j-1}(y_j) = \frac{1}{e} < \infty.
\end{equation}

Because the negative part is integrable, the integral of $f(\lambda)$ is well-defined. Combining the parts yields:
\begin{eqnarray}
\IL  \ln \frac{dP_{i,j}}{dP_{i,j-1}} \, d \barP_j  &=& \IL f^+ \, d\barP_j - \IL f^- \, d\barP_j \nonumber \\
&=& \IDj \ln \frac{d \PMDj}{d \Qjp P_{i,j-1}}(y_j) \, d \PMDj   \nonumber \\
&=& \Dkl(\PMDj \Vert \Qjp P_{i,j-1}) \nonumber \\
&=& \Dkl(P_{i,j} \Vert P_{i,j-1}),
\end{eqnarray}
where the last equality follows from by setting $\barP_j = P_{i,j}$. 
\end{proof}
\subsection{Proof of Lemma~\ref{lem: py}}
\label{proof: py}
\begin{proof}
  \begin{eqnarray}
\Dkl(\barP \Vert P_{1,0}) &=& \IL  \, \ln \frac{d \barP}{d P_{1,0}} \, d\barP =  \IL  \, \ln \Big(\frac{d \barP}{d P_{n,r}} \frac{d P_{n,r}}{d P_{1,0}}\Big) \, d\barP 
 \nonumber \\
&=& \IL  \, \ln \Big( \frac{d \barP}{ d P_{n,r}} \cdot  {\prod_{i=1}^n \prod_{j=1}^{j_i}   \frac{d \PMDj}{d \Qjp P_{i,j-1}}(\Qjl)}\Big) \, d\barP  \nonumber \\
&=& \IL  \, \ln \frac{d \barP}{d P_{n,r}} \, d\barP + \sum_{i=1}^{n}\sum_{j=1}^{j_i} \IL  \, \ln \frac{d \PMDj}{d \Qjp P_{i,j-1}}(\Qjl)\, d\barP \nonumber \\
&\stackrel{\eqref{eq: anyP}}{=}&\Dkl(\barP \Vert P_{n,r})  + \sum_{i=1}^{n}\sum_{j=1}^{j_i} \Dkl(P_{i,j}\Vert P_{i,j-1}).
\end{eqnarray}
 \end{proof}
\subsection{proof of Proposition~\ref{prop: basic}}
\begin{proof}
Using Lemma \ref{lem: py}, we have:
\begin{equation}\label{eq: finite}
\sum_{i=1}^{n}\sum_{j=1}^{j_i} \Dkl(P_{i,j}\Vert P_{i,j-1}) <  \infty.
\end{equation}
Since KL-divergence is non-negative, $\Dkl(P_{n,j} \Vert P_{n,j-1}) \to 0$ uniformly w.r.t. $j$, as $n\to \infty$ by the monotone convergence theorem.

By Pinsker's inequality, $\Vert P_{n,j} - P_{n,j-1} \Vert_{TV} \leq \sqrt{2 \Dkl(P_{n,j} \Vert P_{n,j-1})}$. Thus, the total variation norm of $P_{n,j} - P_{n,j-1}$ also goes to $0$. Applying the triangle inequality for the total variation norm, we get $\Vert P_{n,j} - P_{n,k} \Vert_{TV} \to 0 $ uniformly with respect to $j,k$, as $n\to \infty$.

2. By the definition of the total variation norm,
\begin{eqnarray}
\Vert \Qjp P_{n,k} - \PMDj \Vert_{TV} &=& \Vert \Qjp P_{n,k} - \Qjp P_{n,j} \Vert_{TV} \nonumber \\
&=& 2 \sup_{B \subseteq \MD_j} \Big\vert P_{n,k}(Q_j^{-1}(B)) - P_{n,j}(Q_j^{-1}(B)) \Big\vert \nonumber \\
&\leq& 2 \sup_{A \subseteq \Lambda} \Big\vert P_{n,k}(A) - P_{n,j}(A) \Big\vert \nonumber \\
&=& \Vert P_{n,k} - P_{n,j} \Vert_{TV}.
\end{eqnarray}
Since $\Vert P_{n,k} - P_{n,j} \Vert_{TV} \to 0$ uniformly as $n \to \infty$ from the first statement, it follows that $\Vert \Qjp P_{n,k} - \PMDj \Vert_{TV} \to 0$ uniformly.
\end{proof}

\subsection{Proof of Lemma~\ref{lem: limequal}}
\label{proof: limequal}
\begin{proof}
Without loss of generality, we only consider the case when $r=t$. Following the notation defined in \eqref{eq: formula} we have 
\begin{eqnarray}
 \Dkl(P_{n,t} \Vert P_{1,0}) &=& \int \sum_{i=1}^{n} \sum_{j=1}^t \ln \frac{d \PMDj}{d \Qjp P_{i, j-1}}(Q_j(\lambda)) \, d P_{n,t} \nonumber \\
 &=& \sum_{j=1}^t \int \prod_{k=j+1}^{t} w_{n,k}(\Qkl)  \big(\sum_{i=1}^{n}\ln w_{i,j}(\Qjl)\big) \, d P_{n,j}.
\end{eqnarray}
By the second statement of Proposition \ref{prop: basic}, we have 
$$ \int | w_{n,k}(\Qkl) -1| \, dP_{n,k-1} \to 0,$$
which implies $P_{n,k-1}(|w_{n,k}(\Qkl) -1| >\epsilon ) \to 0$, as $n \to \infty$ for some arbitrary $\epsilon >0$.
From first statement of Proposition \ref{prop: basic}, we have
 $$P_{n,j}(|w_{n,k}(\Qkl) -1| >\epsilon ) \leq P_{n,k-1}(|w_{n,k}(\Qkl) -1| >\epsilon ) + \Vert P_{n,j} - P_{n,k-1} \Vert_{TV}  \to 0. $$
 
Moreover, the boundedness of $w_{i,j}$ and the uniform integrability of $ \sum_{i=1}^n \ln w_{i,j} \subseteq L^1(\PMDj)$ lead to 
 \begin{eqnarray} 
  &&\Dkl(P_{n,t} \Vert P_{1,0}) - \sum_{i=1}^{n}\sum_{j=1}^{t} \Dkl(P_{i,j}\Vert P_{i,j-1})  \nonumber \\
  && \stackrel{\eqref{eq: anyP}}{=} \sum_{j=1}^t \int \big(\prod_{k=j+1}^{t} w_{n,k}(\Qkl)-1\big)  \sum_{i=1}^{n}\ln w_{i,j}(\Qjl) \, d P_{n,j} \to 0,
 \end{eqnarray}
 as $n \to \infty.$
\end{proof} 

\subsection{Proof of Lemma~\ref{lem: wisu.i.}}
\label{proof: wisu.i.}
\begin{proof}
Let $G(x) = x \ln (1+x) $ which is a non-negative increasing function on $[0, +\infty]$ and $\frac{G(x)}{x} \to +\infty, \, x\to \infty$. By de la Vallée-Poussin theorem \cite{Bog07}[Thm 4.5.9] it remains to show 
$$
\sup_{n>1} \IL G(\vert W_{n,r} \vert) \, d P_{1,0} < \infty.
$$
Since 
\begin{eqnarray}
\IL G(\vert W_{n,r} \vert) \, d P_{1,0} &\leq&  \IL (1+  W_{n,r} \ln W_{n,r} ) \, d P_{1,0}, \nonumber \\
&=& 1 + \IL W_{n,r}  \ln W_{n,r} \, d P_{1,0} = 1 + \Dkl(P_{n,r} \Vert P_{1,0}). \nonumber
\end{eqnarray}
Then, it follows from Lemma \ref{lem: limequal} that $W_{n,r}$ is uniformly integrable w.r.t. $P_{1,0}$.
\end{proof}

\subsection{Proof of Theorem~\ref{thm: converge}}
\label{proof: converge}

\begin{proof}
By Dunford–Pettis theorem \cite[Thm 4.7.18]{Bog07}, uniform integrability of $\{W_{n,r}\}_{n,r}$ implies it is also relatively weakly compact in $L^{1}(P_{1,0})$. Therefore, there exists a subsequence that converges weakly to some $\hat{W}$ in $L^1(P_{1,0})$, for which the limiting probability measure is defined as $d\hat{P} = \hat{W} dP_{1,0}$. W.L.O.G., we assume the subsequence to be $\{W_{n_k, t}\}$.

 The weak convergence implies $P_{n_k, t}(Q^{-1}_j (B_j)) \to \hatP(Q^{-1}_j (B_j) ) $ for every measurable set $B_j$. By Proposition \ref{prop: basic}, $\Vert \Qjp P_{n_k,t} - \PMDj \Vert_{TV} \to 0$. Therefore, $\Qjp \hat{P} = \PMDj$ for all $j=1, \dots, t$. Thus, $\hat{P} \in \MC$.

Next, the the functional $W \mapsto  \int W \ln W dP_{1,0}$ is lower semi-continuous w.r.t. weak convergence of $W_{n_k,t}$. Consequently,
\begin{eqnarray}\label{eq: lower_semi}
\Dkl(\hat{P} \Vert P_{1,0}) \leq \liminf_{k \to \infty} \Dkl(P_{n_k, t} \Vert P_{1,0}) = \sum_{i=1}^{\infty}\sum_{j=1}^{t} \Dkl(P_{i,j}\Vert P_{i,j-1}).
\end{eqnarray}

Because $P^* \in \MC$ and $\Dkl(P^* \Vert P_{1,0}) < \infty$, we apply Lemma \ref{lem: py} to $P^*$:
\begin{equation} \label{eq: py*}
\Dkl(P^* \Vert P_{1,0}) = \Dkl(P^* \Vert P_{n,t}) + \sum_{i=1}^{n}\sum_{j=1}^{t} \Dkl(P_{i,j}\Vert P_{i,j-1}).
\end{equation}
 Taking the limit as $n \to \infty$,
\begin{equation}\label{eq: true_bound}
\sum_{i=1}^{\infty}\sum_{j=1}^{t} \Dkl(P_{i,j}\Vert P_{i,j-1}) \leq \Dkl(P^* \Vert P_{1,0}).
\end{equation}

Combining \eqref{eq: lower_semi} and \eqref{eq: true_bound}, we have
\begin{equation}
\Dkl(\hat{P} \Vert P_{1,0}) \leq \Dkl(P^* \Vert P_{1,0}).
\end{equation}
Because $P^*$ is the uniquely minimize the KL divergence over the convex set $\MC$, $\hat{P} = P^*$. Since this holds for every convergent subsequence, the entire sequence $P_{n,t}$ converges weakly to $P^*$. Moreover, this forces \eqref{eq: lower_semi} to be an equality, so
\begin{equation}
\Dkl(P^* \Vert P_{1,0}) = \sum_{i=1}^{\infty}\sum_{j=1}^{t} \Dkl(P_{i,j}\Vert P_{i,j-1}).
\end{equation}
Since by \eqref{eq: py*}  $$ \Dkl(P^* \Vert P_{n,t}) = \Dkl(P^* \Vert P_{1,0})  - \sum_{i=1}^{n}\sum_{j=1}^{t} \Dkl(P_{i,j}\Vert P_{i,j-1}),$$ $\Dkl(P^* \Vert P_{n,t}) \to 0,  n \to \infty$.
Finally, by Pinsker's inequality, the convergence also holds in the total variation norm:
\begin{equation}
\Vert P_{n,t} - P^* \Vert_{TV} \leq \sqrt{2 \Dkl(P^* \Vert P_{n,t})} \to 0.
\end{equation}
\end{proof}

\subsection{Proof of Proposition~\ref{prop: weak-dependence-bounds}}
\label{proof: weak-dependence-bounds}
\begin{proof}
We adopt the notation of $w_{i,j}$ defined in \eqref{eq: formula}. We prove \eqref{eq: wd-conclusions} by induction at the order $(1,1),\ldots,(1,t),(2,1),\ldots, (n,j)$. By assumption~\ref{assp: mismatch}, $\eqref{eq: wd-conclusions}$ holds for $(1,1)$. Now assume this holds for every order before \((n,j)\). 

We first express the sum $\sum_{i=1}^{n}\ln w_{i,j}$ in terms of sums from previous iterations.
At the update $(n,j)$, define the function excluding all $j$-th update by 
\begin{equation}
  h_{n,j}(y_{-j})
  :=
  \sum_{k<j}\sum_{i=1}^{n}\ln w_{i,k}(y_k)
  +
  \sum_{k>j}\sum_{i=1}^{n-1}\ln w_{i,k}(y_k),
\end{equation}
where $\sum_{i=1}^{n=0}:= 0$, \(y_{-j}:=(y_k)_{k\ne j}\) and \(Q_{-j}:=(Q_k)_{k\ne j}\). By \eqref{eq: formula},
\[
  \frac{dP_{n,j-1}}{dP_{1,0}}(\lambda)
  =
  \exp\left\{
    \sum_{i=1}^{n-1}\ln w_{i,j}(Q_j(\lambda))
    +h_{n,j}((Q_{-j}(\lambda)))
  \right\}.
\]
Notice for any $B_j \in \MD_j$,
$$
\int_{Q_j^{-1}(B_j)} \frac{d \Qjp P_{n,j-1}}{d\Qjp P_{1,0}} (Q_j) \, dP_{1,0} = \int_{Q_j^{-1}(B_j)}  \frac{dP_{n,j-1}}{d P_{1,0}} \, dP_{1,0}.
$$
Conditioning under $P_{1,0}$ on $Q_j=y_j$ gives
\begin{align}
  \frac{d\Qjp P_{n,j-1}}{d\Qjp P_{1,0}}(y_j)
  &=
  \exp\left\{\sum_{i=1}^{n-1}\ln w_{i,j}(y_j)\right\}\nonumber \\ 
  &\quad{}\times
  \int e^{h_{n,j}(y_{-j})}
       K_j(y_j,d(y_{-j})).\label{eq: condition}
\end{align}
The conditional integral is finite and bounded away from zero since it only contains sums from previous order.  Multiplying both side of \eqref{eq: condition} by $w_{n,j}=\frac{d\PMDj}{d(\Qjp P_{n,j-1})}$ and taking logarithm yields
\begin{align}
  \sum_{i=1}^{n}\ln w_{i,j}(y_j)
  &=
  \ln\frac{d\PMDj}{d(\Qjp P_{1,0})}(y_j)
  \nonumber\\
  &\quad{}-
  \ln\int e^{h_{n,j}(y_{-j})}
       K_j(y_j,d(y_{-j})).
  \label{eq: wd-update}
\end{align}

 Each of the $t-1$ accumulated sums in
$h_{n,j}$ is essentially bounded and has essential oscillation
at most $B$. Thus $ h_{n,j} \in [a,a+(t-1)B]$ for some finite $a$, $K_{j}(y_j, \cdot)-$a.e. for $\PMDj$-almost every $y_j$. 
We compute
\begin{eqnarray}
  \int e^{h_{n,j}-a} \,dK_j(y_j,\cdot)
  &=& \int_{0}^{\infty}  K_j(y_j,\{e^{h_{n,j}-a}>s\})\,ds  \nonumber \\
  &=& 1+\int_1^{e^{(t-1)B}}K_j(y_j,\{e^{h_{n,j}-a}>s\})\,ds.
\end{eqnarray}
Thus \eqref{eq: wd-kernel} implies
\begin{equation}
  \left|
    \int e^{h_{n,j}-a}  \,dK_j(y_j,\cdot)
    -
    \int e^{h_{n,j}-a} \,dK_j(y_j',\cdot)
  \right|
  \leq\delta\left(e^{(t-1)B}-1\right).
\end{equation}
Both integrals are at least one. Dividing by either integral
and then taking logarithms therefore yields
$$
  \left|
    \ln\int e^{h_{n,j}-a}\,dK_j(y_j,\cdot)
    -
    \ln\int e^{h_{n,j}-a}\,dK_j(y_j',\cdot)
  \right|
  \leq
  \ln\left(1+\delta\left(e^{(t-1)B}-1\right)\right).
$$
Since multiplying the integrand by $e^a$ adds the same constant
to both logarithms, it follows that
\begin{equation}
  \operatorname{osc}_{\PMDj}
  \left(
    y_j\longmapsto\ln\int e^{h_{n,j}}\,dK_j(y_j,\cdot)
  \right)
  \leq
  \ln\left(1+\delta\left(e^{(t-1)B}-1\right)\right).
  \label{eq: wd-conditional-oscillation}
\end{equation}

By \eqref{eq: wd-update}, \eqref{eq: wd-conditional-oscillation},
and \eqref{eq: wd-invariant-bound},
\begin{equation} \label{eq: oscB}
  \operatorname{osc}_{\PMDj}
  \left(\sum_{i=1}^{n}\ln w_{i,j}\right)
  \leq
  \eta+\ln\left(1+\delta\left(e^{(t-1)B}-1\right)\right)
  \leq B.
\end{equation}
Since
$$
  \ln w_{n,j}
  =
  \sum_{i=1}^{n}\ln w_{i,j}
  -
  \sum_{i=1}^{n-1}\ln w_{i,j},
$$
\eqref{eq: oscB} gives
$\operatorname{osc}_{\PMDj}(\ln w_{n,j})\leq2B$.
Furthermore,
\[
  \int_{\MD_j} w_{n,j}\,d(\Qjp P_{n,j-1})=1.
\]
Because $\Qjp P_{n,j-1}$ is a probability measure and is mutually absolutely continuous to $\PMDj$, $$
  \operatorname*{ess\,inf}_{\Qjp P_{n,j-1}}\ln w_{n,j}
  \leq0\leq
  \operatorname*{ess\,sup}_{\Qjp P_{n,j-1}}\ln w_{n,j}.
$$
Combining these inequalities with the oscillation bound yields
$-2B\leq\ln w_{n,j}\leq2B$ almost everywhere and 
$$
  e^{-2B}\leq
  \frac{d\PMDj}{d(\Qjp P_{n,j-1})}
  \leq e^{2B}
  \qquad \PMDj\text{-a.e.},
$$
uniformly in $n$ and $j$. 

Now we show that $\{\sum_{i=1}^{n} w_{i,j}\}_{n}$ is also uniformly integrable in $L^{1}(\PMDj)$ under Assumption~\ref{assp: exist}.
Choose $\barP\in\bigcap_{k=1}^{t}\MC_k$ such that
$\Dkl(\barP\Vert P_{1,0})<\infty$.
By the Pythagorean identity
\eqref{eq: py},
\begin{equation}
  0
  \leq
  \int_{\MD_j}\sum_{i=1}^{n}\ln w_{i,j}\,d\PMDj
  =
  \sum_{i=1}^{n}\Dkl(P_{i,j}\Vert P_{i,j-1})
  \leq \Dkl(\barP\Vert P_{1,0}).
\end{equation}
Therefore, a bounded function with essential oscillation at most $B$
differs from its mean by at most $B$:
$$
  -B\leq\sum_{i=1}^{n}\ln w_{i,j}
  \leq\Dkl(\barP\Vert P_{1,0})+B
  \qquad \PMDj\text{-a.e.},
$$
which shows the uniform integrability  in $L^1(\PMDj)$.
\end{proof}

\end{document}